\documentclass[twocolumn]{aastex63}

\usepackage{natbib}
\usepackage{hyperref}
\usepackage{multirow}
\usepackage{amsfonts,amsmath,amssymb}
\usepackage{graphicx}
\usepackage{color}
\usepackage{xcolor}
\usepackage{tabularx}
\usepackage{enumitem}
\newcommand{\vth}{v_{\text{th}}}
\newcommand{\XSTAR}{\textsc{\scriptsize XSTAR}}
\newcommand{\Athena}{\textsc{Athena\scriptsize ++}}
\newcommand{\etal}{et~al.}

\def\kms{\hbox{km~s$^{-1}$}}

\def\ang{$\text \AA$}

\defcitealias{Dannen2020}{D20}

\begin{document}

\title{ON SYNTHETIC ABSORPTION LINE PROFILES OF THERMALLY DRIVEN WINDS FROM ACTIVE GALACTIC NUCLEI}

\correspondingauthor{Shalini Ganguly}
\email{ganguly@unlv.nevada.edu}

\author[0000-0002-8256-5982]{Shalini Ganguly}
\author[0000-0002-6336-5125]{Daniel Proga}
\affiliation{Department of Physics \& Astronomy, 
University of Nevada, Las Vegas \\
4505 S. Maryland Pkwy, 
Las Vegas, NV, 89154-4002, USA}
\author[0000-0002-5205-9472]{Tim Waters}
\affiliation{Department of Physics \& Astronomy, 
University of Nevada, Las Vegas \\
4505 S. Maryland Pkwy, 
Las Vegas, NV, 89154-4002, USA}
\affiliation{Theoretical Division, Los Alamos National Laboratory}
\author[0000-0002-5160-8716]{Randall C. Dannen}
\affiliation{Department of Physics \& Astronomy, 
University of Nevada, Las Vegas \\
4505 S. Maryland Pkwy, 
Las Vegas, NV, 89154-4002, USA}
\author[0000-0002-1954-8864]{Sergei Dyda}
\affiliation{Institute of Astronomy, Madingley Road, Cambridge CB3 0HA, UK}

\author[0000-0002-1329-658X]{Margherita Giustini}
\affiliation{Centro de Astrobiología (CSIC-INTA), Camino Bajo del Castillo s/n, Villanueva de la Cañada, 28692 Madrid, Spain}

\author[0000-0002-5779-6906]{Timothy Kallman}
\affiliation{NASA Goddard Space Flight Center, Greenbelt, MD 20771, USA}

\author[0000-0002-7868-1622]{John Raymond}
\affiliation{Harvard-Smithsonian Center for Astrophysics, Cambridge, MA, USA}
\affiliation{University of Wisconsin-Madison, Madison, WI, US}

\author{Jon Miller}
\affiliation{Department of Astronomy, University of Michigan, Ann Arbor, MI}

\author[0000-0003-0677-785X]{Paola Rodriguez Hidalgo}
\affiliation{University of Washington Bothell, Bothell, WA, US}

\begin{abstract}
The warm absorbers observed in more than half of all nearby active galactic nuclei (AGN) 
are tracers of ionized outflows located at parsec scale distances from the central engine.
If the smallest inferred ionization parameters correspond to plasma at a few $10^4$~K,
then the gas undergoes a transition from being bound to unbound provided it is further heated 
to $\sim 10^6$~K at larger radii. Dannen \etal{} recently discovered that under these circumstances,
thermally driven wind solutions are unsteady and even show very dense clumps due to thermal 
instability. To explore the observational consequences of these new wind solutions, 
we compute line profiles based on the one-dimensional simulations of Dannen \etal.
We show how the line profiles from even a simple steady state wind solution 
depend on the ionization energy (IE) of absorbing ions, which is a reflection 
of the wind ionization stratification. 
To organize the diversity of the line shapes, we group them into four categories: weak Gaussians, saturated boxy profiles with and without an extended blue wing, and broad weak profiles. The lines with profiles in the last two categories are produced by ions with the highest IE 
that probe the fastest regions. Their maximum blueshifts agree with the highest flow velocities 
in thermally unstable models, both steady state and clumpy versions. In contrast, the maximum blueshifts of the most high IE lines in thermally stable models can be less than half of the actual solution velocities.
Clumpy solutions can additionally imprint distinguishable absorption troughs at widely separated velocities.
\end{abstract}


\section{Introduction} \label{sec:intro}

Seyfert galaxies generally display absorption lines with relatively small blueshifts ($\sim$~100\,--1000 \kms) in their UV and X-ray spectra, indicating the presence of mass outflows \citep{Reynolds1997,Crenshaw1999}. In some such galaxies, the X-ray absorbers (the so-called warm absorbers, WAs) and UV absorbers have very similar velocities which suggests that these absorbers 
are nearly cospatial. This, in turn, suggests that regions of very different temperatures coexist and that the absorption occurs in a multi-phase outflow \citep[e.g.,][and references therein]{Shields1997,Crenshaw1999, Gabel2003, Longinotti2013, Ebrero2016, Fu2017, Mihdipour2017}. 

Mass outflows with relatively small velocities such as these could be thermally driven winds launched at large distances from 
the central engine of an active galactic nuleus (AGN).
There are a few plausible origins for these distant outflows: 
(1) a parsec-scale Compton-heated disk wind \citep[e.g.,][]{Begelman1983, Woods1996, Waters2021};
(2) a wind blown from the inner regions of a torus of dusty material \citep[e.g.][]{Balsara1993,Doro2008,Doro2008b,Doro2012,Doro2016,Kallman2019}; (3) an outflow driven directly from gas infalling towards an SMBH \citep[e.g.,][]{Proga2007, Proga2008,Kurosawa2009, Kurosawa2009b}; and (4) a quasi-spherical outflow driven from an even simpler structure such as a static (non-rotationally supported) 
shell of gas \cite[e.g.,][]{Everett2007}. 

A basic challenge faced by any theoretical model is whether it can account for both the velocity and ionization range inferred from observations. The calculations by \cite{Sim2012}, who computed spectra based on the initial, exploratory simulations from \cite{Kurosawa2009}, illustrate the difficulties typically encountered. They found that while the outflow velocities 
are roughly consistent with observations, the column density and ionization state of the outflowing gas are too high in the models. As such, type (3) models are unlikely to account for WAs, which are often observed in AGNs when there is sufficient sensitivity to detect them \citep[][]{Reynolds1997,Blustin2005, Ricci2017, McKernan2007, Laha2014}.

\cite{Sim2012} discussed a few changes to type (3) models that could
reduce the discrepancy between the model predictions and the observations. While a global change of the gas density or X-ray flux is always a possibility, they noted that it might be more promising to have smaller scale changes in the structure of the outflowing gas.  For a similar total mass outflow rate, higher density might be achieved if it were more strongly clumped. Such a change would not only lead to lower ionization but could also reduce the column density because smaller clumps would subtend smaller solid angles as seen by the central source. Similar conclusions were later reached by \citet{Mizu2019} in the case of type (1) models. Very recently, \cite{Waters2021} showed that multi-phase versions of type (1) solutions do exist. They found that sight lines probing clumps can indeed provide the necessary low ionization gas column density, while their synthetic line profile calculations showed interesting differences between high ionization and low ionization X-ray lines.  

The work of \cite{Waters2021} represents our latest effort to understand how X-ray irradiation 
can trigger thermal instability \citep[TI,][]{Field1965} in \textit{global} dynamical flows.  
Once TI operates, the denser, cooler gas inferred to be necessary to account for the observed 
ionization states of  WAs can naturally be produced. 
The first breakthrough in this line of modeling work came while modeling irradiated inflows. 
Namely, it was shown how the in situ production of multi-phase gas can be triggered using
global time-dependent hydrodynamical simulations of Bondi-like accretion flows
\citep[e.g.,][]{Proga2008,Kurosawa2009, Kurosawa2009b,Barai2012, Moscibrodzka2013, Gaspari2013,  Takeuchi2013}. 
In an outflow regime akin to type (4) models, \citet[hereafter \citetalias{Dannen2020}]{Dannen2020} recently demonstrated that thermally driven outflow solutions can also become clumpy due to TI for a certain range of parameters.

In this paper, we explore the observational consequences of this new type (4) model, 
which is an easily reproducible example of a multi-phase thermally driven radial AGN outflow. 
Specifically, here we present line profile calculations for global simulations that capture 
the dynamics of smooth flows as well as clumpy outflows as they expand and transition back 
into a smooth flow. Our aim here is two-fold:
(i) understanding line profiles from 1D spherically symmetric solutions before 
performing similar studies for multi-dimensional models and 
(ii) assessing if these simple solutions are themselves a promising model of WAs.
Our first aim is related to checking if simple 1D models produce simple 
line profiles for all ions or whether they produce a complex and diverse class of profiles.

This paper is organized as follows. 
In section~\ref{sec:models}, we present all the models analysed in this work. 
In section~\ref{sec:method}, we summarize our methods 
of calculating synthetic line profiles for these models. 
In section~\ref{sec:res} and~\ref{sec:disc}, we present the results of our line profile calculations and our discussion.

\begin{deluxetable*}{cccccccc}[ht]
\tablenum{1}
\tablewidth{0pt}
\tablecaption{Model parameters and some gross properties of steady wind models A21, B and C.
These models share the following basic parameters: $\Gamma=0.3$, $\Xi_0 = 3.12$ ($\xi_0 = 5.0$~erg cm s$^{-1}$) and $M_{\rm BH}=10^6M_\odot$.}
\tablehead{\colhead{Model} & \colhead{HEP$_0$} & \colhead{$r_0$} & \colhead{$\rho_0$} & \colhead{$v$} & \colhead{$\dot{M}$} & \colhead{$r_{\rm out}$} & \colhead{$N_H$}\\
 &  & \colhead{[$10^{18}$ cm]} & \colhead{[$10^{-18}$ g cm$^{-3}$]} &
\colhead{[km s$^{-1}$]} & \colhead{[$10^{24}$ g s$^{-1}$]} & \colhead{[$10^{18}$ cm]} & \colhead{[$10^{23}$ cm$^{-2}$]} }
\startdata
A21 & 14.7 & 0.91 & 3.28 & 659.3 & 1.1 & 9.14 & 1.22 \\
B & 11.9 & 1.13 & 2.15 & 346.0 & 3.4 & 11.28 & 1.47 \\
C & 9.1 & 1.48 & 1.25 & 248.0 & 6.1 & 14.76 & 1.35
\enddata
\tablecomments{D20 found that models B and C are unsteady. 
However, \cite{Waters2021} refined
the numerical methods and found that these two models can eventually settle down to an unstable
yet steady state. Once these solutions are perturbed they become clumpy in agreement
with \citetalias{Dannen2020}'s findings.
The second, third, and fourth columns list the hydrodynamic escape parameter $\rm{HEP}_0$, the inner radius $r_0$, and the density $\rho_0$ at $r_0$, respectively. The next three columns list the following properties of the solutions:
the flow velocity $v$ and  mass loss rate $\dot{M}$ at the outer most grid point, $r_{\rm out}$, at a late time when the flow has become steady. The last column shows the hydrogen column density $N_H$ for the outflow.}
\label{tab:steady}
\end{deluxetable*}

\section{Models} \label{sec:models}

Using the magnetohydrodynamics code \Athena{} \citep{Stone2020}, \citetalias{Dannen2020}
solved the equations 
of non-adiabatic gas dynamics with a radiation force due to electron scattering 
for an unobscured AGN SED \citep[that of NGC 5548; see][]{Mehdipour2015}. The line force 
in these warm/hot winds can exceed gravity but it is much weaker than the gas pressure force 
\citep{Dannen2019}, and has therefore been ignored in 
\citetalias{Dannen2020} for simplicity. They explored 
a parameter space comprised of the three dimensionless parameters governing thermal wind 
solutions: 
(1) Eddington fraction, $\Gamma$ (=$L/L_{\rm Edd}$ for a system with luminosity $L$), 
which sets the radiation force per unit volume due to electron scattering 
as $F_{\rm rad} = GM_{\rm BH}\rho\,\Gamma/r^2$ ($M_{\rm BH}$ is the black hole mass 
and $\rho$ the gas density at radius $r$), 
(2) the pressure ionization parameter at the base of the wind, $r_0$,
\begin{align}
    \Xi_0 = \frac{p_{\rm rad}}{p}\biggr\rvert_{r=r_0} = \frac{\xi_0}{4\pi c k_B T_0} \label{eq:Xi},
\end{align}
where $p_{\rm rad} = F_X/c = f_X\Gamma L_{\rm Edd}/(4\pi r^2 c)$ is the radiation pressure, 
$f_X \equiv L_X/L$ is the ionizing portion of the SED 
($f_X = 0.36$ for the NGC 5548 SED used by \citetalias{Dannen2020}), 
and $\xi_0  = L_X/(n_0 r_0^2)$ is the density ionization parameter 
(with $n_0 = \rho_0/\bar{m}$, $\bar{m}$ being the mean molecular mass),
and (3) the hydrodynamic escape parameter evaluated at $r_0$, 
$\rm{HEP}_0$. This parameter is defined as the ratio of effective 
gravitational potential and thermal energy at $r_0$,
\begin{align}
    \rm{HEP}_0 \equiv \frac{G M_{\rm BH} (1-\Gamma)}{r_0 c_{\rm s,0}^2} , \label{eq:HEP0},
\end{align}
where $c_{\rm s,0} = \sqrt{\gamma k T_0/\bar{m}}$ is
the adiabatic sound speed of the gas with the temperature, $T=T_0$, and $\gamma$ is adiabatic index assumed to be 5/3. 

To specify the lower boundary conditions of the hydrodynamical simulations 
(i.e., $T_0$, $r_0$, and $\rho_0$), \citetalias{Dannen2020} took the following 
five steps:
(i) selected which SED to use along with its associated `S-curve' and $f_{X}$,
(ii) chose values of $M_{\rm BH}$, $\Gamma$, and $\xi_0$,
(iii) computed  $T_0$ for the value of $\xi_0$ and assuming that
the  gas is in thermal equilibrium, i.e., $T_0= T_{\rm gas} = T_{\rm eq}(\xi_0)$, 
(iv) used $T_0$ to compute $c_{s,0}$
and then used eq.~\ref{eq:HEP0} to obtain the value of $r_0$, 
and finally (v) computed $\rho_0$ using the expression for $\xi_0$.
Keeping the parameters $M_{\rm BH}$, $\Gamma$ and $\xi_0$ fixed, \citetalias{Dannen2020} 
altered  only the value of the HEP$_0$ between the models, which required changing $r_0$ 
and $\rho_0$. Note that once $\xi_0$ is fixed, $\Xi_0$ is fixed as well.
In Table~\ref{tab:steady}, we summarize the parameters of three simulations 
and the gross properties of the corresponding outflow solutions.

An extensive exploration of the parameter space led \citetalias{Dannen2020}
to arrive at a new category of `clumpy' thermal 
wind models occurring at sufficiently large radius (small $\rm{HEP}_0$), namely, solutions 
that are unstable to TI.  As further shown by \cite{Waters2021}, these solutions also reach 
a steady state and only occur beyond a characteristic `unbound' radius given by
\begin{align}
R_{\rm u} = \frac{2}{\gamma} \frac{\gamma-1}{\gamma + 1} \frac{T_C}{T_{\rm c,max}} \,R_{\rm IC}\,(1-\Gamma) \approx 140 \,R_{\rm IC}\,(1-\Gamma),
\label{eq:Ru}
\end{align}
where the second expression is evaluated for $\gamma = 5/3$ and for the
thermal equilibrium or `S-curve' corresponding to \citealt[][]{Mehdipour2015}'s SED. 
The ratio $T_C/T_{\rm c,max} = 4.68\times 10^2$, where $T_C$ is the Compton temperature ($\sim 10^8$~K), $T_{\rm c,max} = T(\Xi_{\rm c,max})$ ($\Xi_{c,max}$ is the critical $\Xi$ for which the gas enters the lower TI zone), and the Compton radius is $R_{\rm IC} = (10^{18}/T_{\rm C}) (M/M_\odot)$~cm. 

\begin{figure}[!h]
    \centering
    \includegraphics[scale=0.7]{ 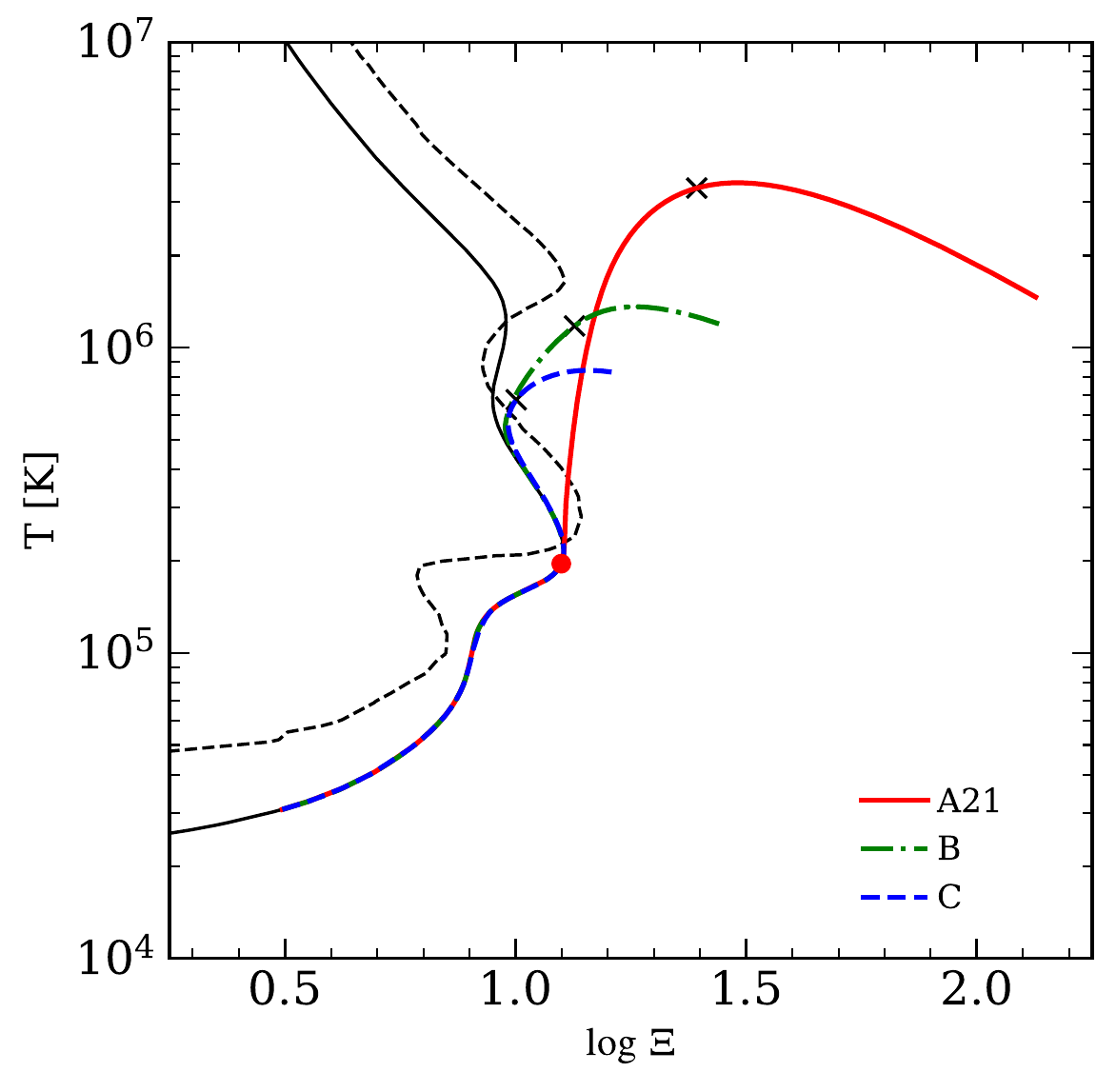}
    \caption{Phase diagram ($T-\Xi$): the solid black line represents the thermal equilibrium 
    curve (S-curve ) while the black dashed line represents the Balbus contour. All points to the left 
    of the latter are thermally unstable whereas those to the right are thermally stable according 
    to Balbus’ criterion for TI \citep{Balbus1986}. There are two unstable regions encountered while 
    tracing the S-curve starting from low temperatures; we refer to them as the lower and upper TI zones. 
    The figure also shows our three steady wind solutions (solid red, dot-dashed green, and dashed blue line 
    for model A21, B, and C, respectively). We marked the sonic points for each solution with an `$\times$'.  
    Notice that upon reaching $\Xi_{\rm c,max}$ (marked by red dot on the S-curve), Model A21 exhibits runaway 
    heating and quickly exits 
    the lower TI zone.  Models B and C instead follow the S-curve within this zone. The decrease in temperature  
    beyond the sonic points in all three cases is due to efficient adiabatic cooling being balanced by radiative heating.}
    \label{fig:ps}
\end{figure}

In Fig.~\ref{fig:ps}, we show this S-curve (solid black line) along with the Balbus contour (black dashed line) that is necessary to analyze TI in a dynamical flow.  
Specifically, while regions of negative slope on the S-curve are thermally unstable,
the Balbus contour maps out the thermally unstable parameter space 
both on and off the S-curve, as explained in the caption.  
These two $T$-$\xi$ relations are computed from our grid of photoionization models first presented in \cite{Dyda2017}.  In their calculations, \cite{Dyda2017} used 
version~2.35 of \XSTAR{} \citep{Kallman2001} and assumed a constant density of 
$n_{\textrm{\tiny XSTAR}}=10^8 \, {\rm cm}^{-3}$ over 
the parameter space 
with $\log\xi$ in the range -2\,--\,8, and $T$ in the range $5\times 10^3$ -- $5 \times 10^8$~K.
We note that, unless otherwise stated, we use cgs units throughout the paper, except for the velocity 
and distance $(r-r_0)$, which are in \kms{} and the inner radius of the computational domain $r_0$, respectively. 

In this figure, we also present examples of steady state solutions (see the legend for color coding of the curves). 
The wind solutions chosen are similar to those presented in 
\citetalias{Dannen2020}:
model A21 is an example of a thermally stable solution\footnote{Model A21 is for 
$\textrm{HEP}_0 = 21$ whereas model A in \citetalias{Dannen2020} is for $\textrm{HEP}_0 = 19$. 
As found by \cite{Waters2021}'s improved numerical methods, the transition
between stable and unsteady solutions occurs at somewhat higher $\textrm{HEP}_0$.}, 
whereas models B and C are examples of thermally unstable solutions.
Although all three models reach a time-independent state, 
they have different properties that signal their stable or unstable character. 
Namely, after they enter the lower TI zone,
the stable solution undergoes rapid heating and leaves the TI zone, evolving
under nearly isobaric conditions (moves vertically up on the phase diagram), 
whereas the unstable solutions continue to follow the S-curve even within 
the lower TI zone (where the slope of the S-curve is negative) 
over a relatively wide range of $T$ before they too leave the zone.

\begin{figure}[!h]
    \centering
    \includegraphics[scale=0.62]{ 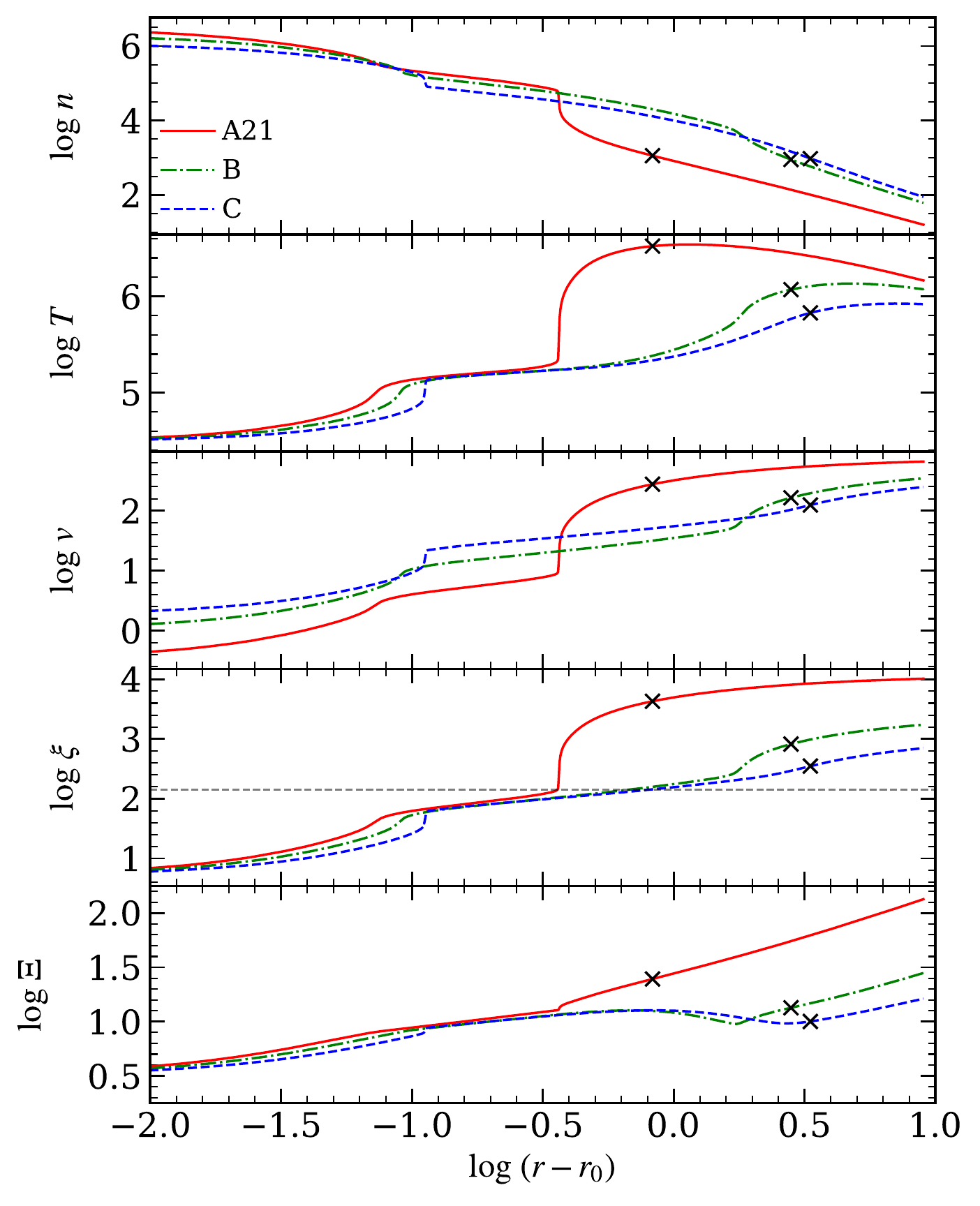}
    \caption{Radial profiles of logarithm of number density $n$ (cm$^{-3}$), temperature $T$ (K), 
    flow velocity $v$ (\kms{}), ionization parameter $\xi$ (erg cm s$^{-1}$) and the dimensionless 
    pressure ionization parameter $\Xi$, for the  three steady wind  models shown in Fig.~\ref{fig:ps}. 
    Sonic points are again marked by `$\times$'s. 
    The horizontal line in the fourth panel marks $\log(\xi_{\rm c,max}) = 2.15$, the entry into the lower TI zone.}
    \label{fig:rad}
\end{figure}

In Fig.~\ref{fig:rad}, we show radial profiles of the main wind properties 
of our three steady state solutions.
Specifically, we plot the number density $n$, temperature $T$, flow velocity $v$, 
ionization parameter $\xi$ and the pressure ionization parameter $\Xi$. 
As discussed above and in \citetalias{Dannen2020}, 
model A21 has quite different properties than models B and C. The main reason for this is that
model A21 undergoes strong runaway heating (as seen by comparing the temperature profiles),
and consequently its velocity/density is higher/lower than that of the other two.
In model A21, $\xi$ is also higher than in models B and C. Therefore we can expect that
the models will produce spectral lines with different profiles.
In general, we also see that $\xi$ increases outward, for all the models. 
One of our goals is to check how this ionization stratification could be traced
in the profiles of absorption lines.

\subsection{Clumpy models}
\label{sec:clumpy_models}
Our steady state model A21 
is formally stable to TI despite it passing through the lower TI zone in Fig.~\ref{fig:ps}.
As discussed by \cite{Waters2021}, this happens whenever there is a sharp 
jump in the entropy profile, which in these solutions will be accompanied by a change in sign 
of the Bernoulli function.  Models B and C, meanwhile, have smooth entropy profiles 
and the Bernoulli function does not change signs when gas encounters the TI zone.  These solutions 
reach a steady state only when there are no more entropy modes being generated at the base of the flow.  
Starting from \citetalias{Dannen2020}'s initial conditions, such entropy modes are only produced 
for a transient period of time. 

\begin{figure}[!h]
    \centering
    \includegraphics[scale=0.6]{ 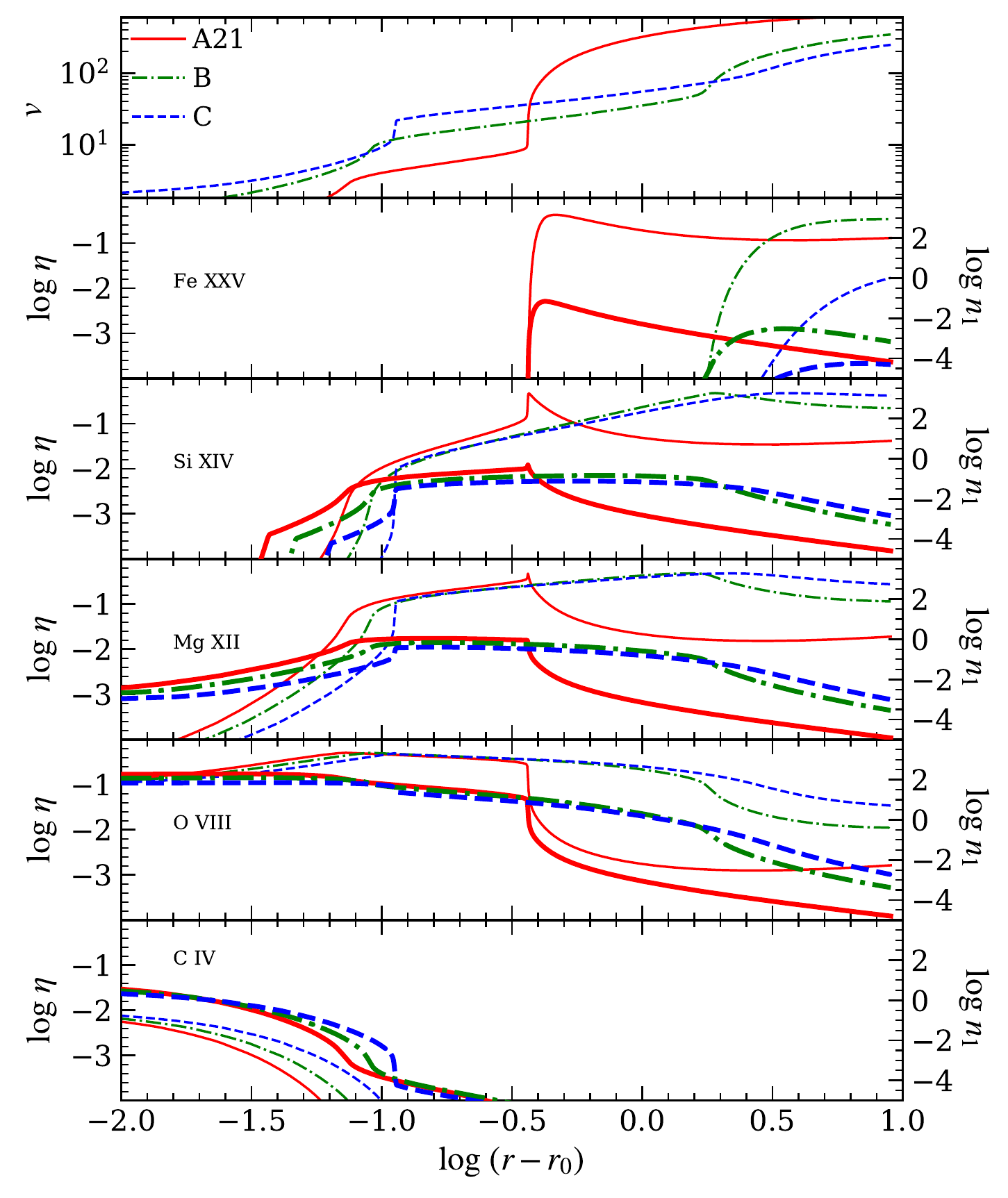}
    \caption{Fractional ionic abundance $\eta$ (shown on left y-axis, with thinner lines) and the ion ground state level population $n_1 = n \eta A$ (shown on right y-axis, with thicker lines) vs $\log \ (r-r_0)$, for the same five ions whose line profiles are compared in Fig.~\ref{fig:all5}. The topmost panel shows radial velocity profiles (in $\rm km\,s^{-1}$) to aid in assessing blueshifts. The legend in this panel applies to both the left and right y-axes.}
    \label{fig:rabund}
\end{figure}

To examine line profiles for clumpy outflow solutions, we reintroduce entropy modes into models 
B and C to trigger TI. Specifically, isobaric perturbations, i.e. density perturbations applied 
at constant pressure (and velocity), are added to the wind base by changing the density boundary 
condition from 
$\rho_{\rm bc} = \rho_0$ to 
\begin{align}
    \rho_{\rm bc}(t) = \rho_0 \left[ 1+\displaystyle\sum_{m=1}^{m=5}A_m\sin\Big(\frac{2\pi t}{T_m}\Big)\right], \label{eq:bcm5}
\end{align}
where $A_m = A_0/N_{\rm modes}$ and $T_m=mT_0$ are the amplitude and period of the driving modes; 
we set $A_0=0.01$ and $T_0=t_{dyn}/20$ (where $t_{\rm dyn} = |r/v|$ is the dynamical timescale). 
We chose this form of density perturbation, comprising five driving modes ($N_{\rm modes}=5)$, 
as it adequately captures the overall complexity of a clumpy wind solution in a controlled way.  
We will refer to the perturbed versions of Models~B and C as Models~B-c and C-c. 

\begin{figure}[!h]
    \centering
    \includegraphics[scale=0.6]{ 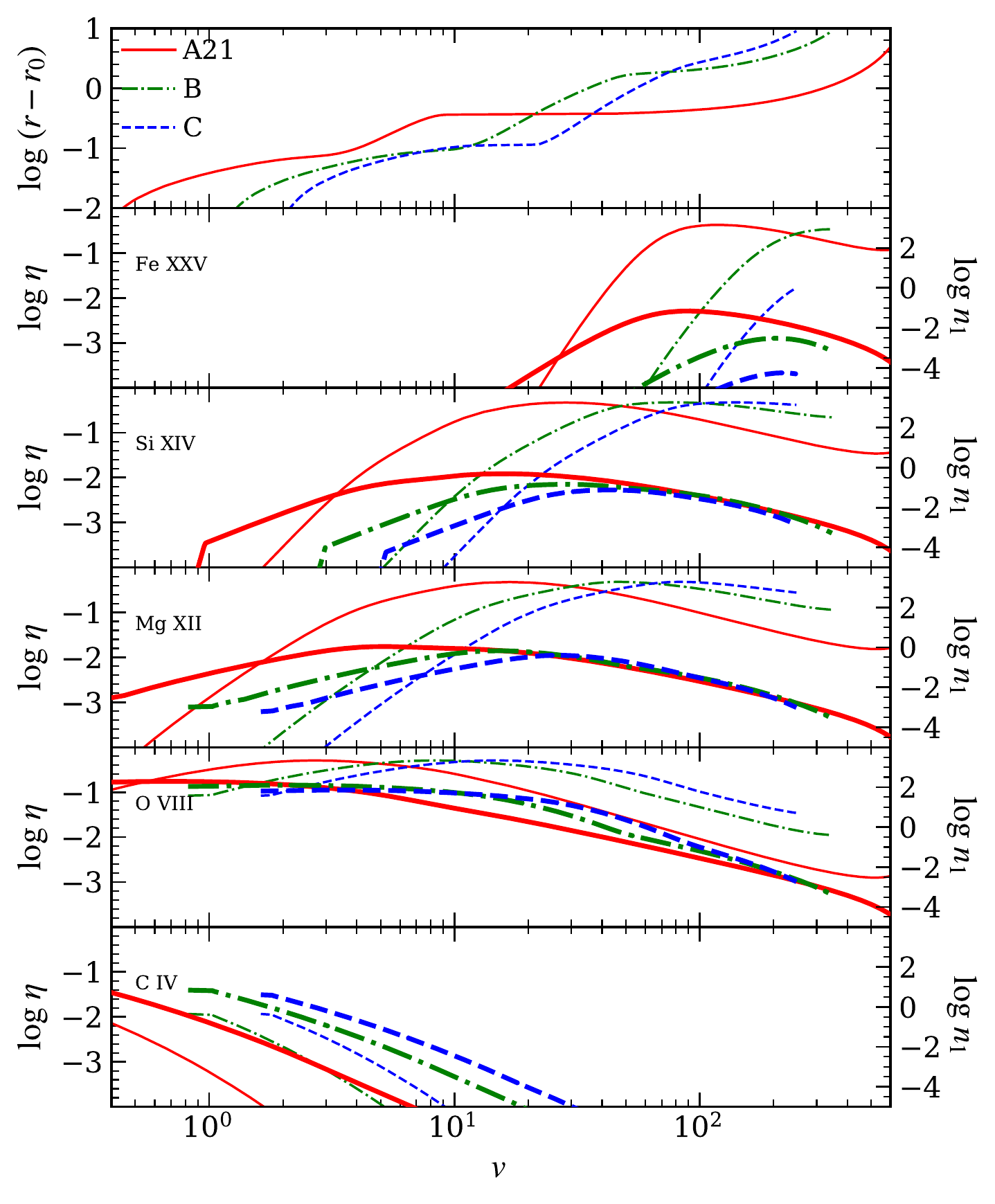}
    \caption{The same quantities in Fig~\ref{fig:rabund} but plotted against the flow velocity $v$ (in $\rm km\,s^{-1}$) rather than distance. The topmost panel shows the radial distribution of $v$.}
    \label{fig:vabund}
\end{figure}

\subsection{Ionization structure and absorption diagnostics}
As an aid to understanding our line profile calculations, we examine 
two other wind properties that enter these calculations, namely the fractional ion abundance, 
$\eta$, and the level population of the ion's 
ground state, $n_1 = nA\eta$ (where $A$ is the elemental abundance relative to hydrogen).
These quantities provide important information about what
to expect and/or how to explain the line profile shapes and the presence of the lines themselves.
We extract $\eta$ and $n_1$ as a part of post-processing 
from our grid of photoionization calculations from \XSTAR{}. These quantities are related to 
the line center absorption coefficient contained in XSTAR output, 
\begin{align}
    (\kappa_{\nu_0}\rho)_{\textrm{\tiny XSTAR}} =& \frac{\pi e^2}{m_e c} n_1 f_{12} \phi(\nu_0) , \nonumber \\
    =& \frac{1}{\sqrt{\pi}} \left[ \frac{\pi e^2}{m_e c}\frac{n_{\textrm{\tiny XSTAR}}A \eta}{\nu_0 (\vth/c)} f_{12} \right] ,\label{eq:kappa}
\end{align}
where $\kappa_{\nu_0}$ is the line center opacity, $\rho = \mu m_p n$ is the gas density (mean molecular mass $\mu=0.6$ for this work, $m_p$ is the proton mass), $f_{12}$ is the oscillator strength of the lower level of the line, $\phi(\nu)$ is the profile function to be defined later, and $v_{\rm th}(r) =\sqrt{2k_B T(r)/m_i}$ is the mean thermal velocity of an ion with mass $m_i$.

To illustrate the details of the ionization structure, 
in Figs.~\ref{fig:rabund} and \ref{fig:vabund}, we plot $n_1$ for five
ions in order of decreasing IE:
\ion{Fe}{25}, \ion{Si}{14}, \ion{Mg}{12}, \ion{O}{8} and \ion{C}{4}
(plotted on the right y-axis using thicker lines in the second to sixth panels). 
We also plot $\eta$ along the left y-axis (shown by thinner lines). 
These two figures are complementary to each other since 
Fig.~\ref{fig:rabund} shows $n_1$ and $\eta$ as functions of radius while 
Fig.~\ref{fig:vabund} shows these quantities as functions of velocity.
This mapping to the velocity space is done solely for the purpose of aiding 
in our analysis and interpretation of line profile results. 
To this end, the top panels of Fig.~\ref{fig:rabund} and Fig.~\ref{fig:vabund}
show respectively the velocity as a function of distance 
and the distance as a function of velocity.

The ionization stratification is on full display in these figures, e.g.,
we see that \ion{C}{4} is only abundant at the wind base whereas \ion{Fe}{25} 
only at very large radii. Meanwhile, \ion{Si}{14} and \ion{Mg}{12} are present at
intermediate locations and show up roughly where the \ion{C}{4} abundance drops off. 
Comparing with Fig.~\ref{fig:ps} or Fig.~\ref{fig:rad}, we see that \ion{C}{4} 
disappears at $r > r(\xi_{\rm c, max})$. 
This is not merely a property of the models but rather the underlying cause explaining why $\xi_{\rm c, max}$ identifies a bend in the S-curve: \ion{C}{4} and other mildly ionized elements provide efficient cooling via line emission, and as these ions rapidly deplete as $\xi$ increases, 
the photoionization equilibrium temperature must sharply increase as well in order to compensate for this lack of efficient cooling.
As we will describe below, this enhanced heating and subsequent gas acceleration results 
in a drop in the amount of matter (measured by the column density). 

\begin{figure}[!h]
    \centering
    \includegraphics[scale=0.65]{ 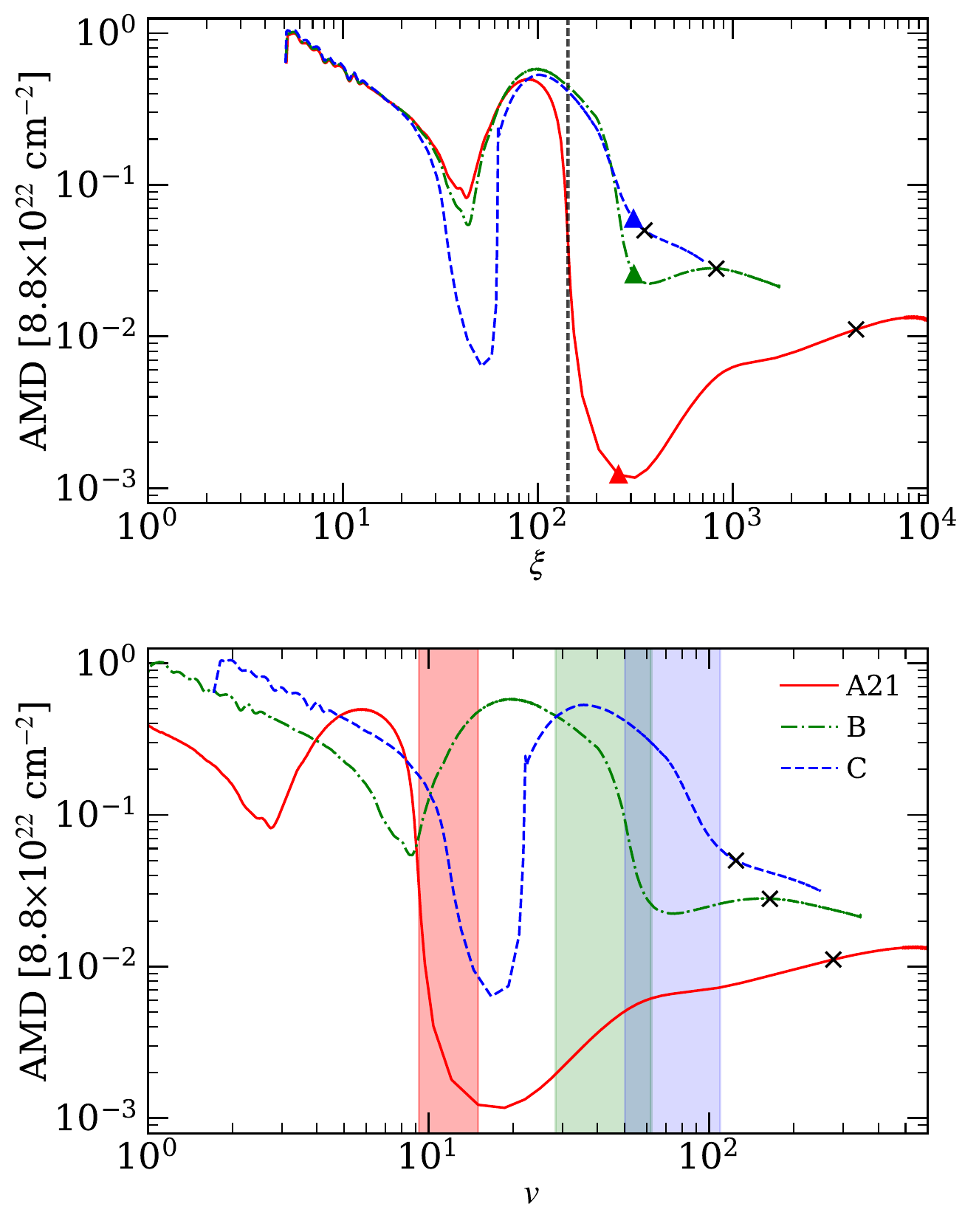}
    \caption{Comparison of the AMD for our three steady state wind solutions.
    The top panel shows the AMD (eq.~\ref{eq:AMD}) in units of its
    maximum value for model A21 (=$8.8\times 10^{22} {\rm cm}^{-2}$) versus $\xi$.  
    The vertical black dotted line marks the base of the lower TI zone, $\xi=\xi_{c, max}$, while the three triangles mark the points where each model leaves this TI zone.
    In the bottom panel, these AMDs are mapped onto the flow velocity, $v$ (in \kms{}; the range with $v <1~$\kms{} is not shown).  Sonic points are marked with `$\times$' symbols and the color shadings denote the parameter space of gas occupying the lower TI zone.}
    \label{fig:amd}
\end{figure}

Our last diagnostic that can help to develop a feel for the absorption properties
of our wind solution is the absorption measure distribution (AMD). It is 
defined as
\begin{align}\label{eq:AMD}
    {\rm AMD} = \frac{dN_H}{d(\log \xi)} = n\lambda_\xi,
\end{align}
where $N_H$ is the hydrogen column density and $\lambda_\xi =(d\log{\xi}/dr)^{-1}$ 
is the characteristic length scale for variations in $\xi$. The AMD has proven to be both 
an important observable \citep{Holczer2007,Behar2009} and a useful tool for quantifying 
the absorption properties of hydrodynamical models. We note that for a steady state, $v\propto \xi$ ($v\propto1/nr^2$, as is $\xi$). Therefore, $\lambda_{\xi}=\lambda_v \equiv (d\log{v}/dr)^{-1}$.  

In Fig.~\ref{fig:amd}, we show the AMDs for our three steady models.  
There are two prominent dips in the AMD: one at the entrance to the TI zone, which is marked by the vertical line in the top panel, and another at even smaller $\xi$, corresponding to gas on the 
cold phase branch of the S-curve.  The bottom panel in Fig.~\ref{fig:amd} shows this latter dip occurs for low velocity gas.
As pointed out by \cite{Dyda2017}, a dip in the AMD can be caused by enhanced heating even 
in regions that are thermally stable. Referring to Fig.~\ref{fig:ps}, notice that the cold branch 
of the S-curve becomes nearly vertical 
(for $\log\Xi \approx 0.9$ and $T$ just above $10^5$~K),
or equivalently, the S-curve and the Balbus contours approach each other.  
In this physical regime, the velocity is relatively low but the acceleration is relatively high 
(e.g., see Fig.~\ref{fig:rad} for the steep gradient in the radial profiles of $T$, $n$ and $v$). 
This rapid acceleration leads to the minima in the AMD 
at $v\approx 2$, 10, and 20~\kms{} in models A21, B, and C, respectively. 

A quantitative understanding of the reason for a decrease in the AMD is found 
by referring to the right hand side of eq.~\ref{eq:AMD} and noting that 
$\lambda_{\xi}=\lambda_v \equiv (d\log{v}/dr)^{-1}$.  
Also note that an enhanced acceleration results in shortening
the characteristic length scale $\lambda_v$.  The region of the enhanced acceleration
is relatively small. Therefore, $\lambda_v$ increases as the slope of the S-curve 
flattens again, and the AMD recovers to the level from before the drop. 
Importantly, our analysis above (mentioned while describing Figs.~\ref{fig:rabund} and~\ref{fig:vabund}) reveals that such a change in the slope of the S-curve can be driven by a reduction in line cooling: 
all wind solutions reach this region of enhanced acceleration where the abundance of \ion{C}{4} undergoes a sharp decrease (at $\log(r-r_0) \approx -1$ in the bottom panel of Fig.~\ref{fig:rabund}).
In other words, the gas experiences enhanced heating as a result of it losing important coolants.

Farther downstream, the gas becomes even more ionized and, as we already mentioned,
it enters the lower TI zone at $\xi_{\rm c,max}$. The corresponding changes in the thermal 
and dynamical properties of the wind solutions produce a change in the AMD. Namely,
the AMD drops by one or even two orders of magnitude (see the shaded color areas 
in the bottom panel of Fig.~\ref{fig:amd}).
This is best exemplified by the results for model A21 where 
the heating is so enhanced that it leads to runaway heating. The drops in the AMD 
for models B and C are not as large simply because the heating is weaker.
Upon exiting the TI zone, the AMD in model C continues to decrease unlike in models A21 and B 
where we see a mild increase.

Our analysis demonstrates how the flow properties closely depend on the gas dynamics 
and thermodynamics even if the flow is time independent. The methodology used 
by \citetalias{Dannen2020} to couple photoionization and hydrodynamical calculations 
provides us with a substantial amount of information.  Not only can we extract the opacities 
necessary to compute absorption line profiles, but we can also examine intermediate quantities 
to explain the physics behind the properties of the profiles and the relations 
between different ionization states.

\section{Methods} \label{sec:method}

The photoionization calculations from \XSTAR{} provide us with the quantity in brackets 
in equation~\eqref{eq:kappa} above, for $n=n_{\textrm{\tiny XSTAR}}=10^8$~cm$^{-3}$. Due to $(\kappa_{\nu_0}\rho)_{\textrm{\tiny XSTAR}}$ being insensitive to $n_{\textrm{\tiny XSTAR}}$ 
in low density regimes, we can calculate the opacities corresponding to our hydrodynamic density 
profiles as follows,
\begin{align} \label{corrected}
    \kappa_{\nu_0}\rho = (\kappa_{\nu_0}\rho)_{\textrm{\tiny XSTAR}} \times \frac{n}{n_{\textrm{\tiny XSTAR}}} .
\end{align}
In practice, we generate lookup tables of $(\kappa_{\nu_0}\rho)_{\textrm{\tiny XSTAR}}$ parameterized 
by $\xi$ and $T$, using bilinear interpolation to access values for the local hydrodynamic values 
of ($\xi$,$T$) in our simulations \citep[see][]{Waters2017}.

\begin{deluxetable*}{ c  c  c  c || c c c c }
\centering
\tablecaption{Ions used for computing line profiles (in decreasing order of IE).}
\tablewidth{700pt}
\tablenum{2}
\tablehead{Lines & \colhead{$\lambda_0$} & \colhead{$\Delta v$} & \multicolumn{1}{c||}{IE} & \colhead{Lines} & \colhead{$\lambda_0$} & \colhead{$\Delta v$} & \colhead{IE} \\
& \colhead{[\ang]} & \colhead{[km s$^{-1}$]} & \multicolumn{1}{c||}{[eV]} & & \colhead{[\ang]} & \colhead{[km s$^{-1}$]} & \colhead{[eV]}  }
\startdata
\ion{Fe}{26} & 1.77802 & 911.0 & 9277.69 & \ion{C}{6} & 33.7342 & 48.0 & 489.99 \\
 & 1.78344 & & & & 33.7396 & & \\ \hline
 \ion{Fe}{25} & 1.8505 & \nodata & 8828.0 & \ion{Ne}{8} & 780.324 & \nodata & 239.10 \\ \hline
 \ion{Ar}{18} & 3.7311 & \nodata & 4426.23 &  \ion{O}{6} & 1031.91 & 1648.0 & 138.12 \\ 
 &  & & & & 1037.61 & & \\ \hline
 \ion{S}{16} 4 & 3.9908 & \nodata & 3494.19 & \ion{N}{5} & 1238.82 & 961.0 & 97.89 \\ 
 &  & & & & 1242.80 & & \\ \hline
 \ion{Si}{14} 6 & 6.18042 & 262.6 & 2673.18  & \ion{C}{4} & 1548.2  & 497.0 & 64.49  \\ 
  & 6.18583 & & & & 1550.77  & & \\ \hline
  \ion{Si}{14} 5 & 5.21681  & 65.6 & 2673.18 & \ion{He}{2} & 303.780 & 5.9 & 54.42  \\
  & 5.21795 & & & & 303.786 & & \\\hline
 \ion{Mg}{12} & 7.10577 & 48.1 & 1962.66 & \ion{S}{4} & 1073.51 & \nodata & 47.22 \\  
 & 7.10691 & & &\\ \hline
 \ion{Ne}{10} & 10.2385 & 32.2 & 1362.20 & \ion{S}{4}$^*$ & 1062.66 & \nodata & 47.22 \\
 & 10.2396 & & & \\ \hline
  \ion{O}{8} 19 & 18.9671 & 85.4 & 871.41 & \ion{Si}{4} & 1393.76 & 1927.0 & 45.14 \\
  & 18.9725 & & & & 1402.77 & &\\ \hline
 \ion{O}{8} 16 & 16.0055 & 22.5 & 871.41 & \ion{Si}{3} & 1206.5 & \nodata & 33.49 \\
   & 16.0067 & & &  &  &  &  \\\hline
 \ion{O}{8} 15 & 15.1760 & 9.9 & 871.41 & \ion{C}{2} & 1334.52 & 268.0 & 24.38  \\ 
  & 15.1765 & & & & 1335.71 & & \\ \hline
  \ion{O}{7} & 21.602 & \nodata & 739.29 &  \ion{Mg}{2} & 2798.75 & \nodata & 15.04  \\  \hline
 \ion{N}{7} & 24.7792 & 65.7 & 667.046 & \\
  & 24.7846 & & & & 
 \enddata
\tablecomments{ We denote different transitions for the same ion with a number, which corresponds 
to the wavelength of the transition, as in \ion{Si}{14} $5$ and \ion{Si}{14} $6$. The $2^{\rm nd}$, 
$3^{\rm rd}$, and $4^{\rm th}$ column denote the rest frame wavelength $\lambda_0$, the velocity 
shift (in \kms{}) of the doublet components $\Delta v= c\,(\lambda_b-\lambda_r)/\lambda_r$  
(where subscripts $r$ and $b$ denote red and blue components) and the ionization energy of the ion, IE.
}
\label{tab:all}
\end{deluxetable*}

Our current analysis is an improvement over \cite{Waters2017} on two major fronts: 
(i) their heating/cooling source term was calculated using analytical fits to earlier 
photoionization calculations by \cite{Blondin1994}, while their opacities were extracted 
from \XSTAR{} using a similar $10\,\rm{keV}$ Bremsstrahlung SED.  For this work, we extract 
abundances, opacities, and heating/cooling rates from \XSTAR{} for a realistic AGN SED; 
(ii) we examine over two dozen lines as opposed to the study of only 
the \ion{O}{8} Ly$\alpha$ doublet (\ion{O}{8} 19) in \citet{Waters2017}, which was determined 
to be the strongest line in their local cloud simulations.
In Table~\ref{tab:all}, we list the basic properties of the lines and the corresponding 
ions selected for analysis. 

The frequency-dependent line optical depth is calculated (using the trapezoid rule) as
\begin{align}
    \tau_\nu = \int_{r_{\rm in}}^{r_{\rm out}} \kappa_\nu(r) \rho(r) \, {\rm d}r ,\label{eq:tau}
\end{align}
where $r_{\rm in}$ and $r_{\rm out}$ are the inner and outer radius of the computational domain and 
\begin{align}\label{eq:kappanu}
    \kappa_\nu(r) = \kappa_{\nu_0}(r) \phi(\nu)/\phi(\nu_D)
\end{align}
is the frequency distribution of opacity corresponding to a particular ion.  Here $\phi(\nu)$ 
is the profile function, taken to be a Gaussian distribution with a thermal line width 
$\Delta\nu_0=\nu_0v_{\rm th}/c$: \begin{align}
    \phi(\nu) = \frac{1}{\sqrt{\pi}} \frac{1}{\Delta\nu_0} \exp\left(\frac{-(\nu-\nu_D)^2}{\Delta\nu_0^2}\right). \label{eq:phi}
\end{align}
This function peaks at frequency $\nu_D(r) = \nu_0[1+v(r)/c]$, corresponding to the line center 
Doppler-shifted to the local wind velocity.

The solution of the radiative transfer equation in 1D, for zero emission, can be expressed as the intensity,
\begin{align}
    I_\nu = I_0 e^{-\tau_\nu(r)} , \label{eq:inu}
\end{align}
where $r$ is the radial distance from the central source (an X-ray corona near the black hole).  
We use equations \eqref{eq:tau} - \eqref{eq:inu} to calculate the synthetic absorption line profiles, 
$I_\nu$, for a chosen ion and model run, taking $I_0=1$.

\begin{deluxetable*}{ c | p{1.6cm} p{1.6cm} p{1.6cm} || c | p{1.6cm} p{1.6cm} p{1.6cm} }
\centering
\tablecaption{Ionic column densities (cm$^{-2}$) in steady models.}
\tablewidth{700pt}
\tablenum{3}
\tablehead{ Ions & \colhead{A21} & \colhead{B} & \multicolumn{1}{c||}{C} & Ions & \colhead{A21} & \colhead{B} & \colhead{C}}
\startdata
\ion{Fe}{26}& 2.42$\times 10^{16}$ & 4.71$\times 10^{15}$ & 5.20$\times 10^{13}$ & \ion{Ne}{8} & 1.59$\times 10^{18}$ & 1.61$\times 10^{18}$ & 1.53$\times 10^{18}$ \\
\ion{Fe}{25} & 1.67$\times 10^{16}$ & 1.35$\times 10^{16}$ & 3.86$\times 10^{14}$ & \ion{O}{6} & 5.93$\times 10^{18}$ & 6.05$\times 10^{18}$ & 6.20$\times 10^{18}$\\
\ion{Ar}{18}& 2.17$\times 10^{15}$ & 5.61$\times 10^{15}$ & 3.38$\times 10^{15}$ & 
  \ion{N}{5} & 2.16$\times 10^{17}$ & 2.21$\times 10^{17}$ & 2.27$\times 10^{17}$  \\
 \ion{S}{16}& 8.82$\times 10^{15}$ & 4.28$\times 10^{16}$ & 3.77$\times 10^{16}$ & 
\ion{C}{4} & 8.86$\times 10^{16}$ & 9.06$\times 10^{16}$ & 9.35$\times 10^{16}$  \\
 \ion{Si}{14} & 6.64$\times 10^{16}$ & 2.88$\times 10^{17}$ & 2.75 $\times 10^{17}$ &  \ion{He}{2} & 1.44$\times 10^{19}$ & 1.48$\times 10^{19}$ & 1.50$\times 10^{19}$  \\
 \ion{Mg}{12} & 2.82$\times 10^{17}$ & 6.68$\times 10^{17}$ & 6.04$\times 10^{17}$ & 
\ion{S}{4} & 12.70$\times 10^{13}$ & 2.77$\times 10^{13}$ & 2.86$\times 10^{13}$  \\
 \ion{Ne}{10} & 2.20$\times 10^{18}$ & 3.56$\times 10^{18}$ & 2.99$\times 10^{18}$ & \ion{Si}{4} & 3.89$\times 10^{12}$ & 4.00$\times 10^{12}$ & 4.13$\times 10^{12}$   \\
 \ion{O}{8} & 2.01$\times 10^{19}$ & 2.34$\times 10^{19}$ & 2.01$\times 10^{19}$ & \ion{Si}{3} & 1.96$\times 10^{11}$ & 2.02$\times 10^{11}$ & 2.09$\times 10^{11}$ \\
 \ion{O}{7} & 3.20$\times 10^{19}$ & 3.26$\times 10^{19}$ & 3.21$\times 10^{19}$ & \ion{C}{2} & 1.56$\times 10^{13}$ & 1.60$\times 10^{13}$ & 1.65$\times 10^{13}$\\
  \ion{N}{7} & 3.43$\times 10^{18}$ & 3.67$\times 10^{18}$ & 3.37$\times 10^{18}$ &  \ion{Mg}{2} & 1.25$\times 10^7$ & 1.29$\times 10^7$ & 1.34$\times 10^7$ \\
 \ion{C}{6} & 1.28$\times 10^{19}$ & 1.33$\times 10^{19}$ & 1.29$\times 10^{19}$ & & & &
\enddata
\tablecomments{The ionic column density for each ion in our survey, $N_{\rm ion}$ = $\int n_1 dr$, where $n_1$ is the level population for the ground state, for the three steady models A21, B and C.}
\label{tab:nion}
\end{deluxetable*}

\begin{figure*}
    \centering
    \includegraphics[width=0.668\textwidth]{ 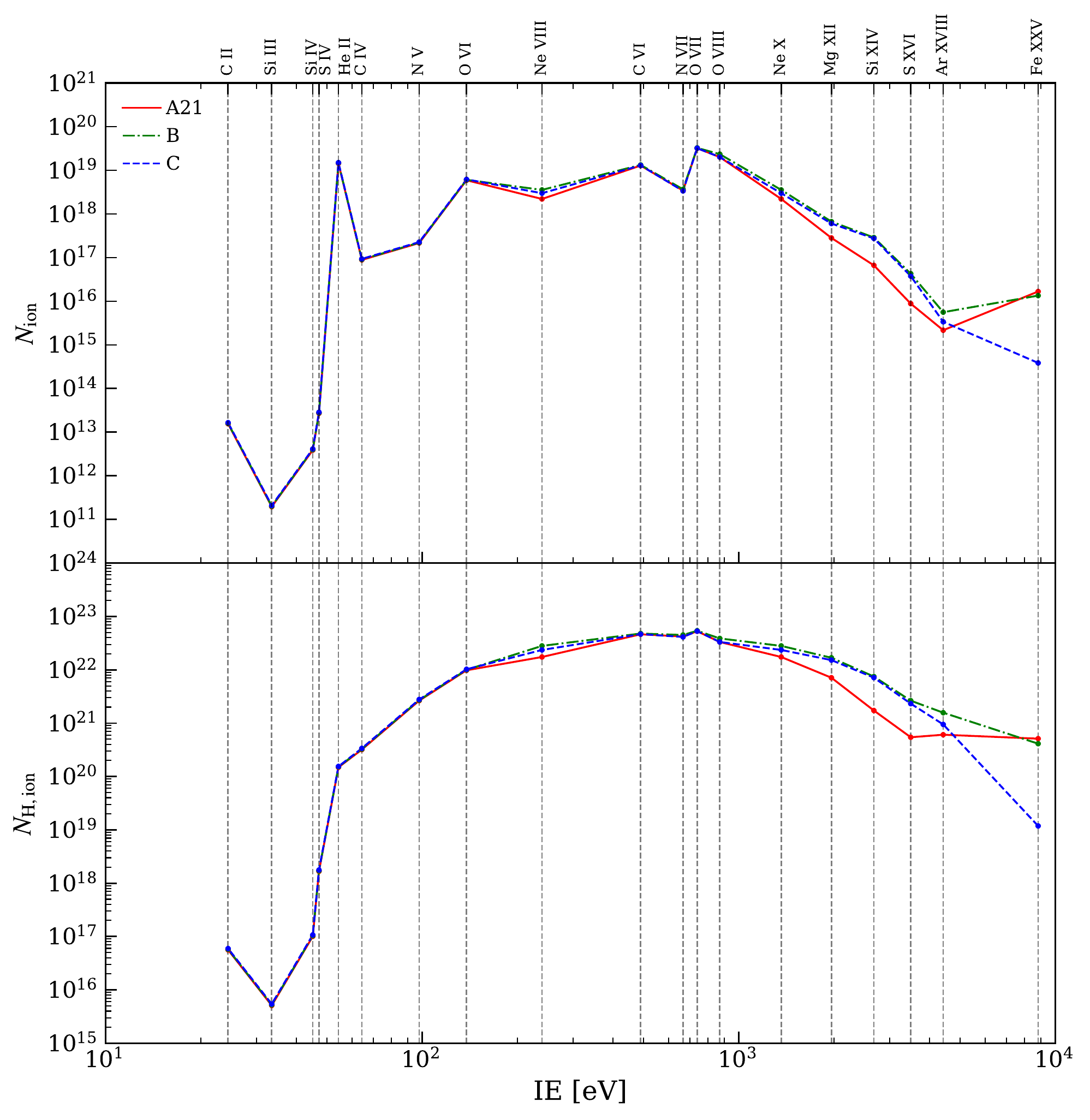}
    \caption{Column densities as functions of ionization energy. 
    {\it Top panel:}  the column
    density of a given ion (the ionic column density, $N_{\rm ion}$) as a function of IE
    of that ion (see also Table~\ref{tab:nion}). {\it Bottom panel:} as above but with $N_{\rm ion}$ corrected for the element abundance ($N_{\rm ion}/A$), so it is the hydrogen column density in the region where a given ion is present.}
    \label{fig:Nion_vs_vB}
\end{figure*}

\section{Results} \label{sec:res}
As one would expect based on the values of the AMD, our models are associated with significant ionic column densities for ions spanning a large range of ionization energy (IE).
In Table~\ref{tab:nion}, we list the results of our calculations of ionic column density 
for the steady models. We also plot these ionic column densities 
and their values corrected for the element abundance, i.e., $N_{ion}/A$, i.e. the hydrogen column density in the region where the given ion is present,
in Fig.~\ref{fig:Nion_vs_vB} (see the top
and bottom panel respectively).  These results show 
that there can be significant absorption in lines due to ions of the abundant elements,
especially if they are  mildly to highly ionized (e.g., \ion{He}{2}, \ion{C}{6},
\ion{O}{7}, and \ion{O}{8}). On the other hand, ions with IE less than that of \ion{He}{2} are expected to produce weak absorption.

In \S{\ref{sec:res_steady}}, we first analyze line profiles computed for steady state solutions.  
Then in \S{\ref{sec:res_clumpy}}, we analyze those of unsteady versions 
of our thermally unstable models (B and C).

\subsection{Steady state wind solutions} \label{sec:res_steady}
\begin{figure}[!h]
    \centering
    \includegraphics[scale=0.65]{ 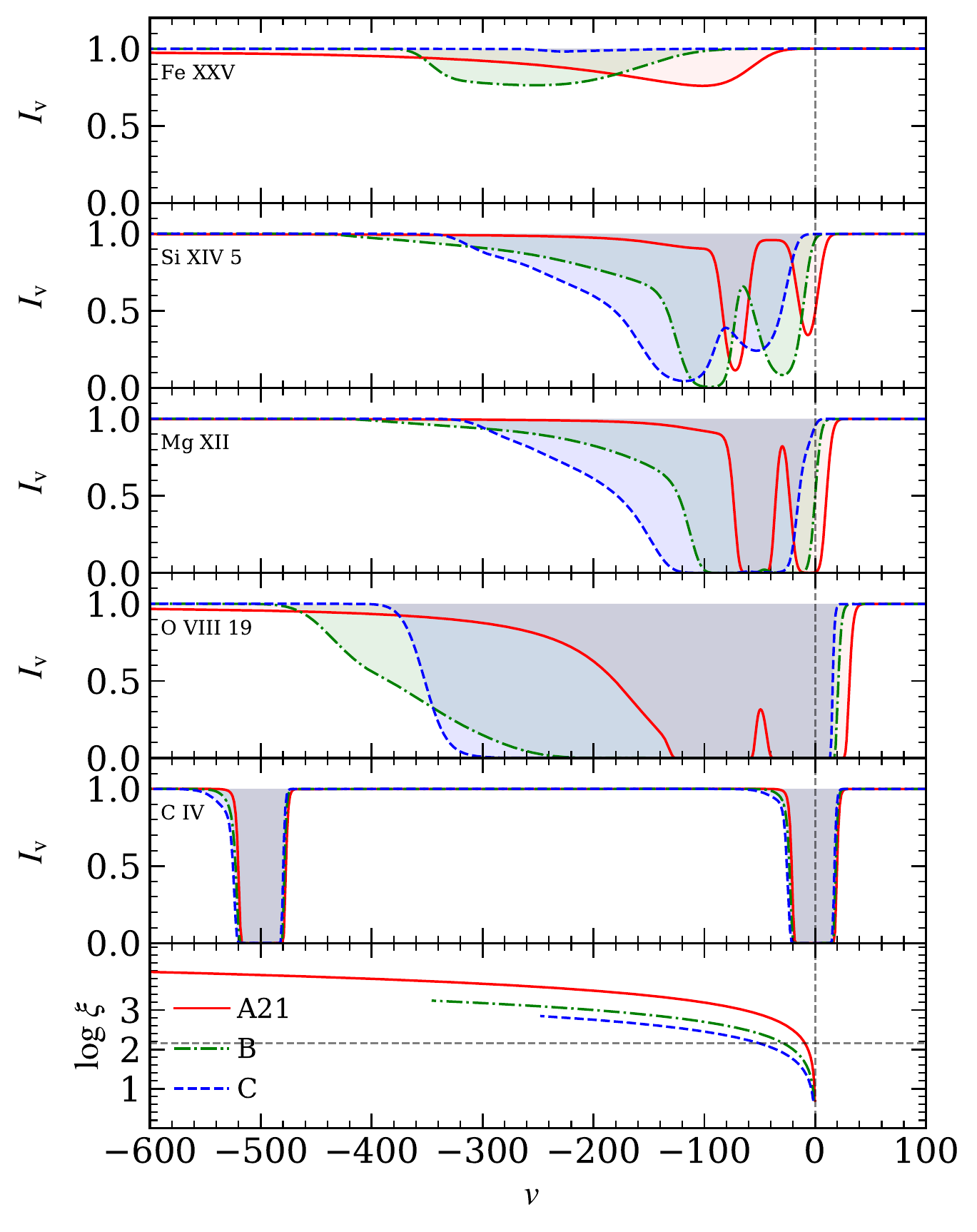}
    \caption{Absorption profiles due to lines from five ions (as labeled) together with a plot of $\xi$ vs. $v$ (bottom panel). 
    Red solid, green dot-dashed, and blue dashed lines denote model A21, B and C, respectively.
    Absorption troughs are shaded to highlight their depth and shape. Except for \ion{Fe}{25}, all lines shown here 
    are doublets (see Table~\ref{tab:all} for details). The \ion{C}{4} doublet components 
    (second from the bottom panel) are separated
    in all three models whereas the \ion{Si}{14}~5, \ion{Mg}{12}, 
    and \ion{O}{8}~19 doublet components are blended. The vertical grey dashed line marks $v=0$ in all panels. 
    In the lowermost panel, the horizontal dashed line marks
    $\log(\xi_{\rm c,max}) = 2.15$ in order to judge the blueshift at which the flow enters the lower TI zone.}
    \label{fig:all5}
\end{figure}

\begin{figure*}[!h]
    \centering
    \includegraphics[height=22cm,width=\textwidth,scale=0.75]{ 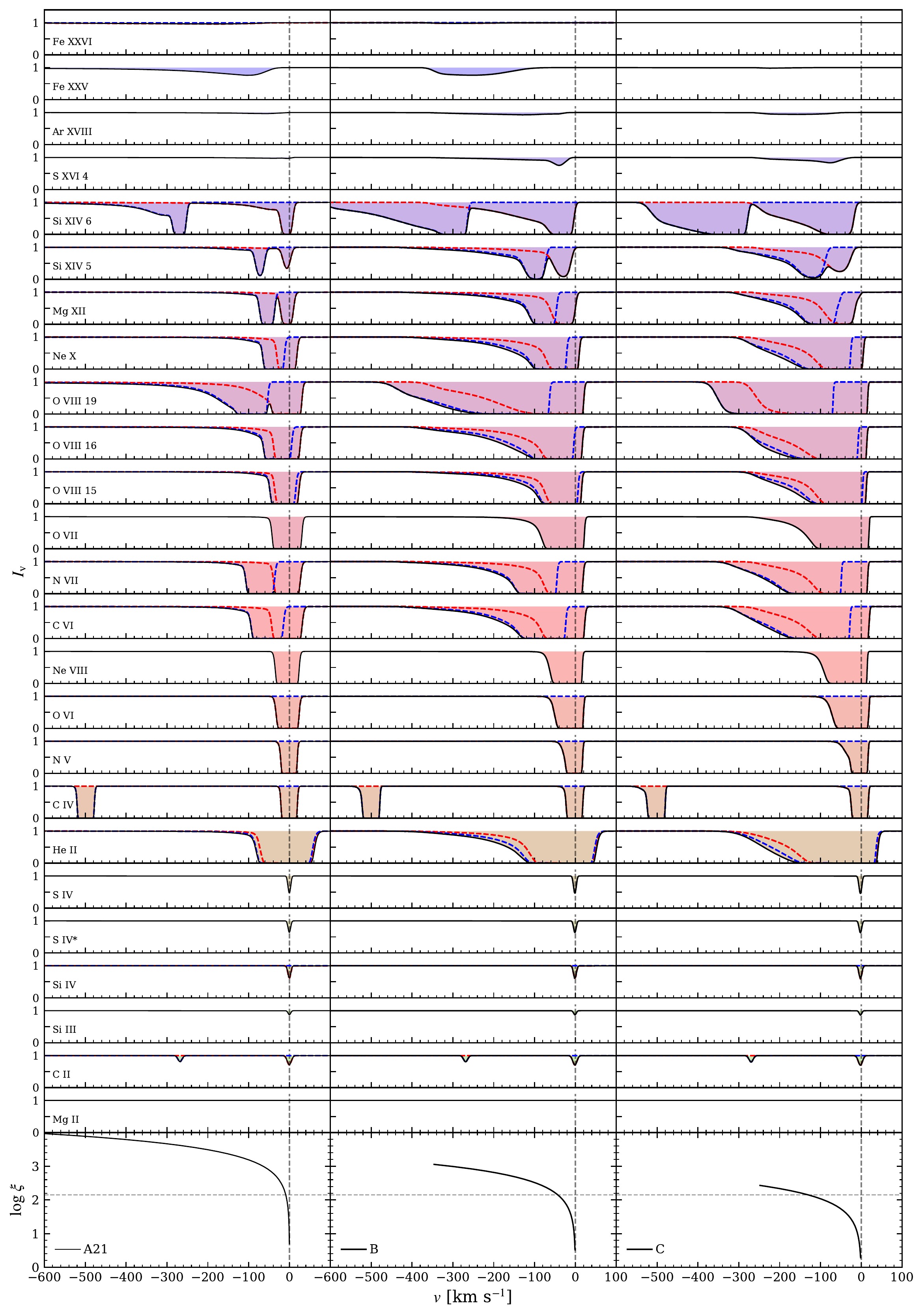}
    
    \caption{Absorption profiles for models A21, B and C (left, middle, and right columns 
    of panels, respectively). See the ion labels at left and Table~\ref{tab:all} for the line identification. 
    The profiles are shown using solid black lines. For doublet lines, we plot a composite 
    spectrum by adding the red and blue components while accounting for their wavelength 
    separation in velocity space. The individual components have been additionally plotted 
    with red and blue dashed lines. In the bottom panel, we plot the ionization parameter 
    against the radial flow velocity, for reference. The dashed horizontal line indicates the value of $\xi_{c, max}$, whereas the black vertical line marks zero velocity.For FITS files of all our line profile calculations, see: \url{http://www.physics.unlv.edu/astro/
    clumpywindsims-lps.html}.}
    \label{fig:A21BC}
\end{figure*}

In Fig.~\ref{fig:all5}, we highlight the absorption profiles for the same ions that we used 
to illustrate the ionization stratification in section~\ref{sec:models} (i.e., \ion{Fe}{25}, \ion{Si}{14}, \ion{Mg}{12}, \ion{O}{8}, 
and \ion{C}{4}).\footnote{The number after the ion name denotes the wavelength (in \ang) 
absorbed by the ion to undergo a particular transition and is only shown for the few lines 
in our full list given in Table~\ref{tab:all} that are from the same ion (e.g., \ion{Si}{14}~5 
and \ion{Si}{14}~6).}
In Fig.~\ref{fig:A21BC}, we show profiles for our full selection of lines for various ions, 
including those shown in Fig.~\ref{fig:all5}. Note that in both the figures, for the doublet lines, 
we plot the total profile by adding the profiles due to the red and blue components. In Fig.~\ref{fig:A21BC}, 
however, we also show the individual profiles due to these components using red and blue dashed lines, 
respectively.
To aid connection between the wind solution and the line profiles, in the bottom panels of these figures, 
we plot $\xi$ as a function of $v$. 

Line profiles in our sample belong to one of four major categories: 
1) Gaussian (typical for weak lines, e.g., the \ion{S}{4} line), 
2) boxy (lines with a strong broad core, 
e.g., the \ion{C}{4} line,
3) boxy with an extend blue wing (e.g., \ion{Mg}{12}), and 
4) weak extended (e.g., \ion{Fe}{25}). 
Below we will focus on the last three categories.

The troughs of the \ion{C}{4}, \ion{O}{8}~19 and \ion{Mg}{12} show 
black saturation. The profile of the red component of the UV \ion{C}{4} line is relatively simple. 
It has a boxy shape and its width is determined by the line saturation. The line has a relatively
sharp transition from the very optically thick core to the line wings.
However, the profile is not exactly symmetric because even though the line forming region is near 
the wind base, it still has
non-zero bulk velocity (see fig.~\ref{fig:rad}). Therefore, the blue wing of the line is broader than
the red wing due to a Doppler blueshift. 
The \ion{C}{4} profile is similar for all three wind 
models because the wind solutions are also similar at the base. 
The main difference is the line width: it increases from model A21 through B to C
and this order is consistent with the order of the velocity at which $\xi=\xi_{c, max}$
(compare the bottom panel in Fig.~\ref{fig:all5} with the panel immediately above).
These subtle differences are in contrast with the other profiles shown in this figure, with different models 
producing quite different line profiles; these are all X-ray lines with significant opacity 
above the the wind base.

The profiles of the \ion{O}{8}~19, \ion{Mg}{12} and \ion{Si}{14}~5
lines are narrower, less blueshifted, and even weaker for model A21 than for models B and C, 
and this might seem contradictory to the fact that the wind in model A21 is the fastest.
However, the lines probe the conditions where the population 
of their lower levels are the largest, and as we showed in the previous 
section, the $n_1$ distribution of their corresponding ions
peaks within a relatively narrow velocity range that does not include the fast part of the wind. 
That is, ionization stratification may cause the signature for any individual line to
be diminished and not reflect on the entire wind solution.
In addition, the strong runaway heating in model A21 leads to a drop in the overall flow density
that in turn reduces wind absorption,
as can be seen in the AMD.  
Therefore, the absorption in these three lines at large velocities in model A21 is weaker than in the other two models.

The \ion{Fe}{25} line stands out among the other lines in Fig.~\ref{fig:all5}:
its blueshifted trough does not extend to zero velocity. This feature is simply 
caused by lack of the \ion{Fe}{25} ion at small velocities (see Figs.~\ref{fig:rabund} 
and \ref{fig:vabund} for the distributions of the ion abundance).

\begin{figure}
    \centering
    \includegraphics[scale=0.8]{ 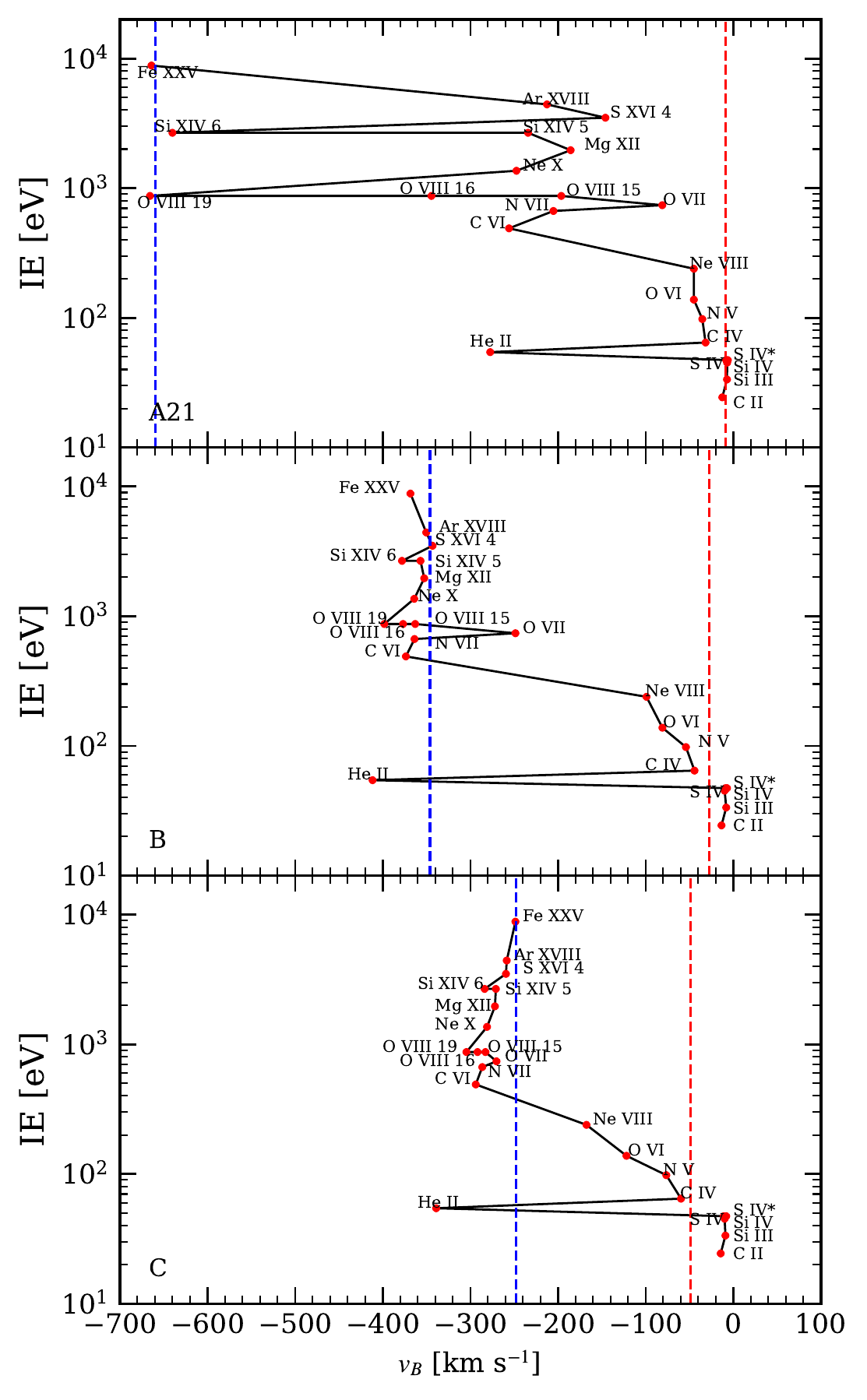}
    \caption{
    The position of the the blue-edge of the absorption trough, $v_B$ 
    as a function of IE. We formally define $v_B$ as the velocity where $I_{\rm v}=0.99$.
    When the line is a blended doublet we correct $v_B$ for the doublet split.
    Here we show results for the steady models. 
    The blue and red dashed lines indicate the maximum flow velocity 
    and the velocity where ionization parameter $\log \xi = \log \xi_{c, max} = 2.15$, respectively.}
    \label{fig:bevsie}
\end{figure}

A quick inspection of Fig.~\ref{fig:A21BC}, where our full sample of line profiles 
are plotted in order of decreasing IE,
reveals that the velocity of the red-edge of the lines increases overall with IE:
it is negative for the low energy ions and can be positive for the high energy ions.
In addition, the extent of the blue wing of strong lines mostly decreases with decreasing IE.
These trends are more clearly evident in Fig.~\ref{fig:bevsie} that will be discussed 
below. The notable exception to the second trend
is the \ion{He}{2} line which is relatively wide. This is due to the fact that the line's strong 
saturation is related to a relatively very high helium abundance leading to
its ground level population being significant over a wide range of radii.

Fig.~\ref{fig:A21BC} shows that a `detachment' of the line forming region 
from the wind base is not unique to the \ion{Fe}{25} line.  
This is seen to a lesser degree in the profiles of the \ion{Si}{14}~5~and~6 lines 
for model C (and lesser still for model B).
Referring yet again to Fig.~\ref{fig:vabund} to check if there is obvious explanation for this,
we see that although models B and C have smaller terminal 
velocities than model~A21, they are actually faster where 
the \ion{Si}{14} ion is most abundant. 

We also note that absorption in the \ion{Mg}{2} line is absent in our models while 
the \ion{C}{2}, \ion{Si}{3}, \ion{Si}{4}, \ion{S}{4} lines are weak and symmetric. 
These low ionization lines form at the wind base and the maximum absorption 
(of the red component) is at or near zero velocity.

To quantify the extent of the blue wing, 
we measured the velocity at which $I_{\rm v}=0.99$ for each line.
We refer to this as the blue-edge velocity, $v_B$,
and we plot $v_B$ as a function of IE in Fig.~\ref{fig:bevsie}.
When the line is a blended doublet, we correct $v_B$ for the doublet split,
while for unblended lines we measure $v_B$ of the red component.

Fig.~\ref{fig:bevsie} shows that, in general, for low IE lines, 
the extent of the blue wing 
is close to $v_{c,{\rm max}}$, the velocity at which $\xi$ approaches $\xi_{c,max}$,
while $v_B$ is close to the maximum wind velocity for high IE lines. 
As mentioned above, the \ion{He}{2} line
is an exception because it is a very strongly saturated line.
We note that in model~A21, not all high IE lines extend to
the maximum wind velocity (see e.g., \ion{Ne}{10} and \ion{Mg}{12}) 
because the opacity of this line is relatively 
small at large radii for this wind solution.

The three lines of the same ion, i.e. those of \ion{O}{8}, offer
a good illustration of how the opacity affects the line extent.
The lines of this ion have very strong and broad cores 
and extended blue wings.
In model A21, $v_B$ increases with
the oscillator strength, which is strongest for \ion{O}{8}~19 and weakest for \ion{O}{8}~15. 
The line absorption extends all the way
to the wind maximum velocity only for the \ion{O}{8}~19 line (see also Fig.~\ref{fig:A21BC}).

Overall, the comparison of the three models shows that $v_B$ of 
high IE lines could be a good proxy for the maximum flow velocity 
for models B and C but for model A21, this is true only for certain high IE lines.

\begin{figure}[!h]
    \includegraphics[scale=0.65]{ 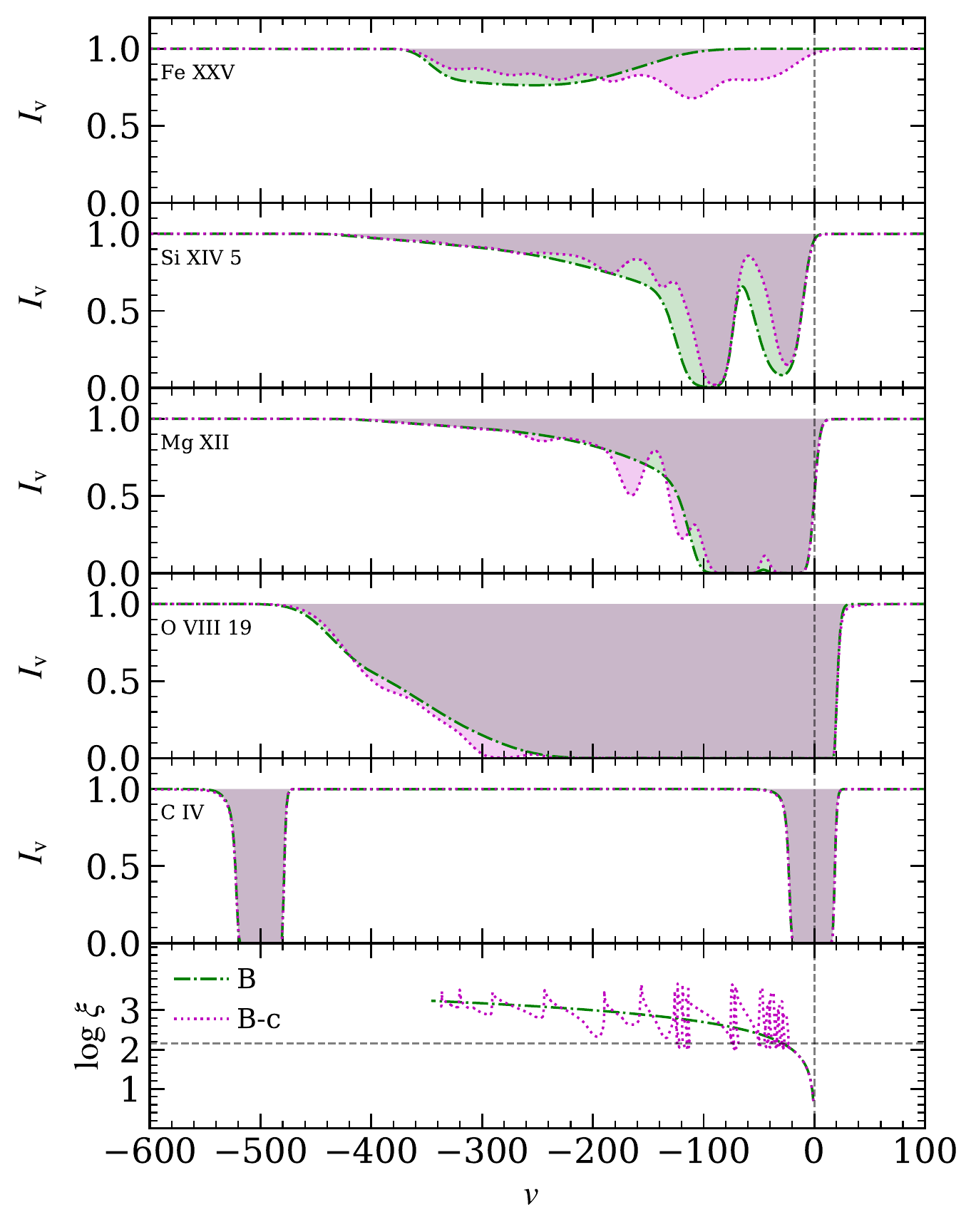}
     \caption{Line profile comparison for the steady model B (dot-dashed green shaded profile) 
     and the unsteady/clumpy model B-c (dotted magenta shaded profile, refer to eq.~\ref{eq:bcm5}). 
     The lower panel shows $\log\xi$ vs $v$ for these two cases, for comparison, where horizontal 
     gray dashed line marks $\log(\xi_{\rm c,max}) = 2.15$, while vertical dashed lines mark $v=0$ 
     in all panels.}
    \label{fig:b5}
\end{figure}

\begin{figure}
    \centering
    \includegraphics[scale=0.65,width=0.47\textwidth]{ 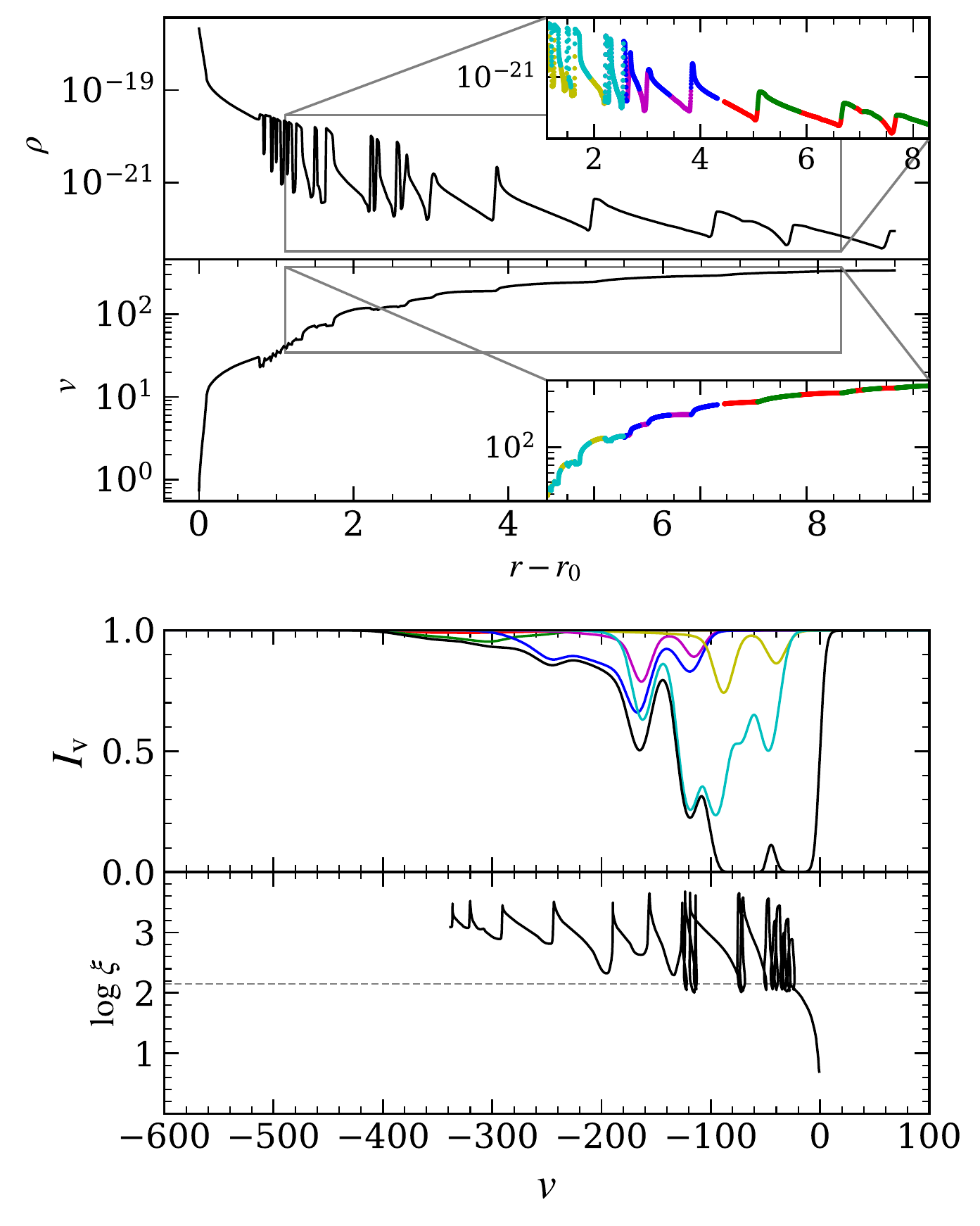}
    \caption{Contributions of different parts of a clumpy outflow (model B-c) to the overall line profile for \ion{Mg}{12}. The colors green, blue and cyan highlight the higher-than-average density regions corresponding to clumps, while the red, magenta and yellow highlight intercloud regions.}
    \label{fig:breakdown}
\end{figure}

\begin{figure}[!h]
    \centering
    \includegraphics[scale=0.65]{ 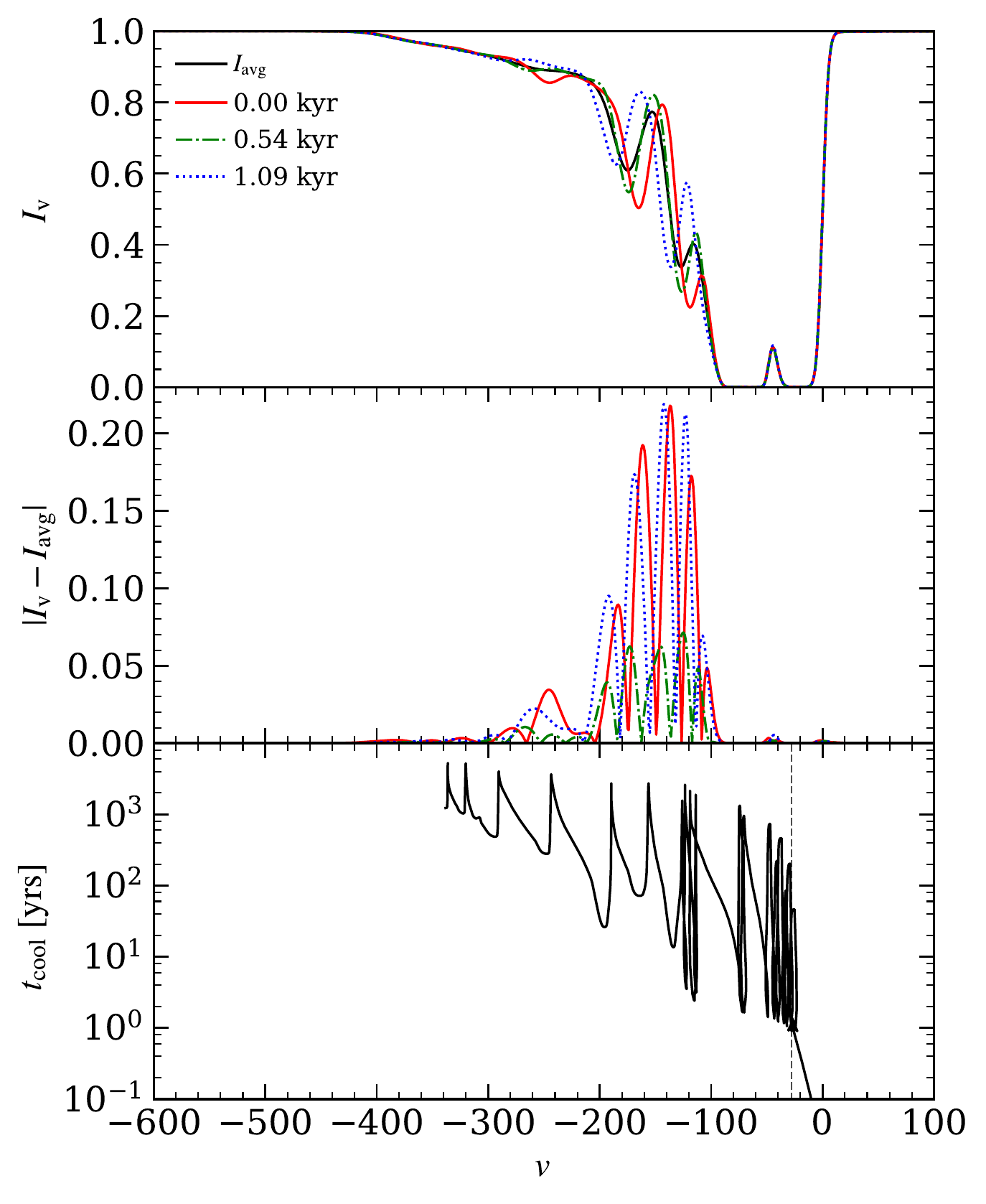}
    \caption{Variability of the \ion{Mg}{12} line profile for model B-c. Line profiles at three different states $t_i$ of the simulation runs are plotted along with the time-averaged line profile.  The legend shows the time $\Delta t = (t_i - t_0)$ (in kyr), the red line denoting the state $t_0$. The middle panel shows the difference between 
    the snapshots from the average profile. The bottom panel shows the cooling time, $t_{\rm cool}$; a vertical line marks the velocity for which $\log \ \xi_{\rm c,max} = 2.15$, showing that cooling times exceed 1 year in the clumps.  Any time-dependence must therefore occur on longer timescales. An animation of the time evolution of line profiles for B-c and C-c, can be found here: \url{http://www.physics.unlv.edu/astro/clumpywindsims-lps.html}}
    \label{fig:avg}
\end{figure}

\begin{figure}
    \centering
    \includegraphics[scale=0.6,width=0.5\textwidth]{ 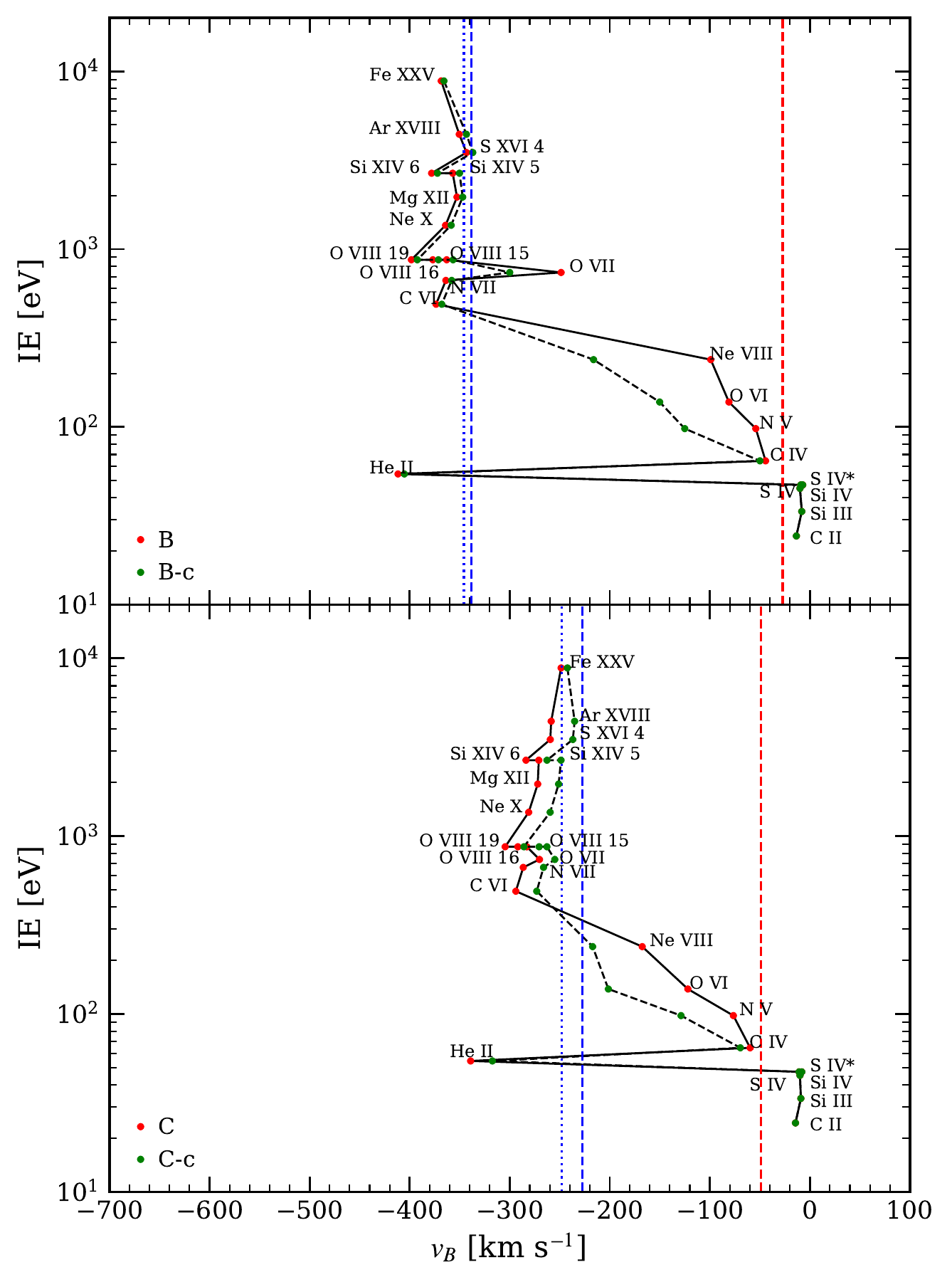}
    \caption{Ionization energy vs velocity of the blue-edge of line profiles (defined as where $I = 0.99$) for steady models (black solid line with red dots) compared to their clumpy versions (black dashed line with green dots). The dotted (dashed) blue and red vertical lines indicate the maximum flow velocity and the velocity where $\xi = \xi_{\rm c,max}$, respectively, for the clumpy (steady) models.}
    \label{fig:bevuie}
\end{figure}

\subsection{Clumpy wind solutions} \label{sec:res_clumpy}
In Fig.~\ref{fig:b5}, we over-plot line profiles for Model~B-c (in magenta) on those for steady Model~B 
shown previously in Fig.~\ref{fig:all5}. 
Only one new category of line profile needs to be added to our list given in \S{\ref{sec:res_steady}}: 
boxy with an extended {\it non-monotonic} blue wing.  
It is clear from this figure that non-monotonic features 
can be due to either stronger or
weaker absorption at different velocities and
are present in all lines except
those from ions with peak abundance
at $\xi < \xi_{\rm c,max}$ like \ion{C}{4}. 
In other words, clumps do not affect the `boxy' profiles characterizing ions formed at the wind base,
as clumps are produced further downstream.

Of the remaining profiles, that of \ion{Mg}{12} is perhaps the most intuitive:
the clumps in Model~B-c can be considered over-densities superimposed on the `background'
wind profiles shown in Fig.~\ref{fig:rad}. Hence, a deeper absorption trough 
should occur at the velocity where the clump resides.  
More distant clumps will thus have both higher velocity and lower density, so this explains the general
trend that absorption troughs get less deep at higher velocities in the wing.  
The width of each trough
is essentially $\max(\Delta v_c, v_{\rm th})$, where $v_{\rm th}$ is the ion thermal velocity
and $\Delta v_c$ is the range of line of sight velocities internal to the clump. As will be shown below, model C-c results in smooth line profiles because the intercloud spacing is very small. Hence, we can state that a very clumpy wind can appear completely smooth with monotonic wings if it satisfies the condition that the velocity separation of the clumps is smaller than the thermal widths.
Note that similar considerations have been used to place constraints on the number of clouds in broad line regions \citep{Arav1997}.  
 
To clearly illustrate the contribution of clumps to line profiles,
Fig.~\ref{fig:breakdown} breaks down different density regions of the outflow in the case of the \ion{Mg}{12} line profile (again for model B-c).  
The saturated part 
of the absorption trough is formed at the wind base, although some of the slowest
clumps do contribute.
This is shown by the cyan colored regions, 
where absorption due to several high velocity clumps blend together 
to form part of the line core as well as discernible absorption troughs in
the blue wing of the profile.
Regions marked in green and blue are responsible for the higher velocity features that together 
add up to form the extended blue wing of the line. 
The absorption within the warm, under-dense regions between the clumps 
is seen to be mostly negligible (see the red, magenta and yellow colors). 

In Fig.~\ref{fig:avg}, we illustrate
how the \ion{Mg}{12} line profile changes
due to the evolution of the clumps in model B-c.
We plot the profile at three times, $t_i$ 
(as explained in figure caption;
$\Delta t = (t_i-t_0)$ = \{0, 0.54, 1.09\} kyr). 
For reference, we also
plot the time-averaged line profile (shown in black).  
The dips become increasingly blueshifted with time and also 
less deep as the clumps 
move outward with a higher velocity and become less dense. 
In the middle panel, we show the absolute difference 
between each of the three profiles 
and the time-averaged profile. 
This reveals that the largest differences occur 
at velocities where the line just starts to de-saturate.  
The bottom panel shows the cooling time,
\begin{align}
    t_{\rm cool} = 6.6 \frac{T_5}{n_4 \mathcal{C}_{23}} \ \textrm{yrs} \label{tcool}
\end{align}
where $T_5 = T/10^5$ K, $n_4 = n/10^4$ cm$^{-3}$ and $\mathcal{C}_{23} = \mathcal{C}/10^{23}$ 
is the cooling rate in units of erg cm$^3$ s$^{-1}$. The vertical dashed line marks the velocity 
at which $\log\ \xi > \log \ \xi_{\rm c,max} = 2.15$. Above this velocity, the cooling time 
is of the order of years, which indicates that the flow cannot respond to minor flux variability on timescales less than $\sim1$ year. 
Note from \S{\ref{sec:clumpy_models}} that we chose the time period of perturbations to be a small 
fraction of the dynamical time scale, but still large enough to ensure that the gas can respond.

\begin{figure*}
    \centering
    \includegraphics[height=22cm,width=\textwidth,scale=0.75]{ 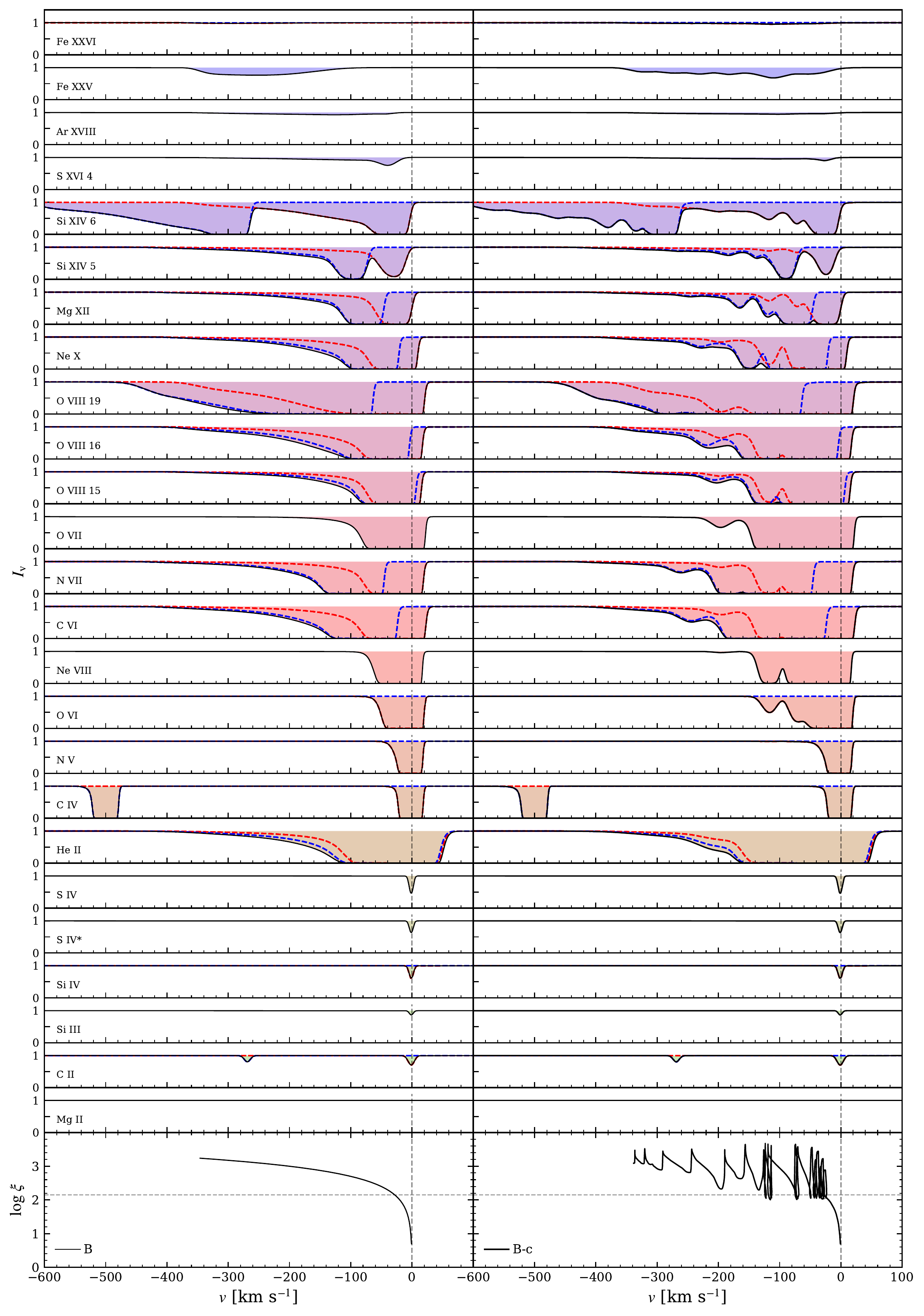}
    \caption{Same as \ref{fig:A21BC}, but for steady model B compared to clumpy model B-c.}
    \label{fig:Bperturb}
\end{figure*}

\begin{figure*}
    \centering
    \includegraphics[height=22cm,width=\textwidth,scale=0.75]{ 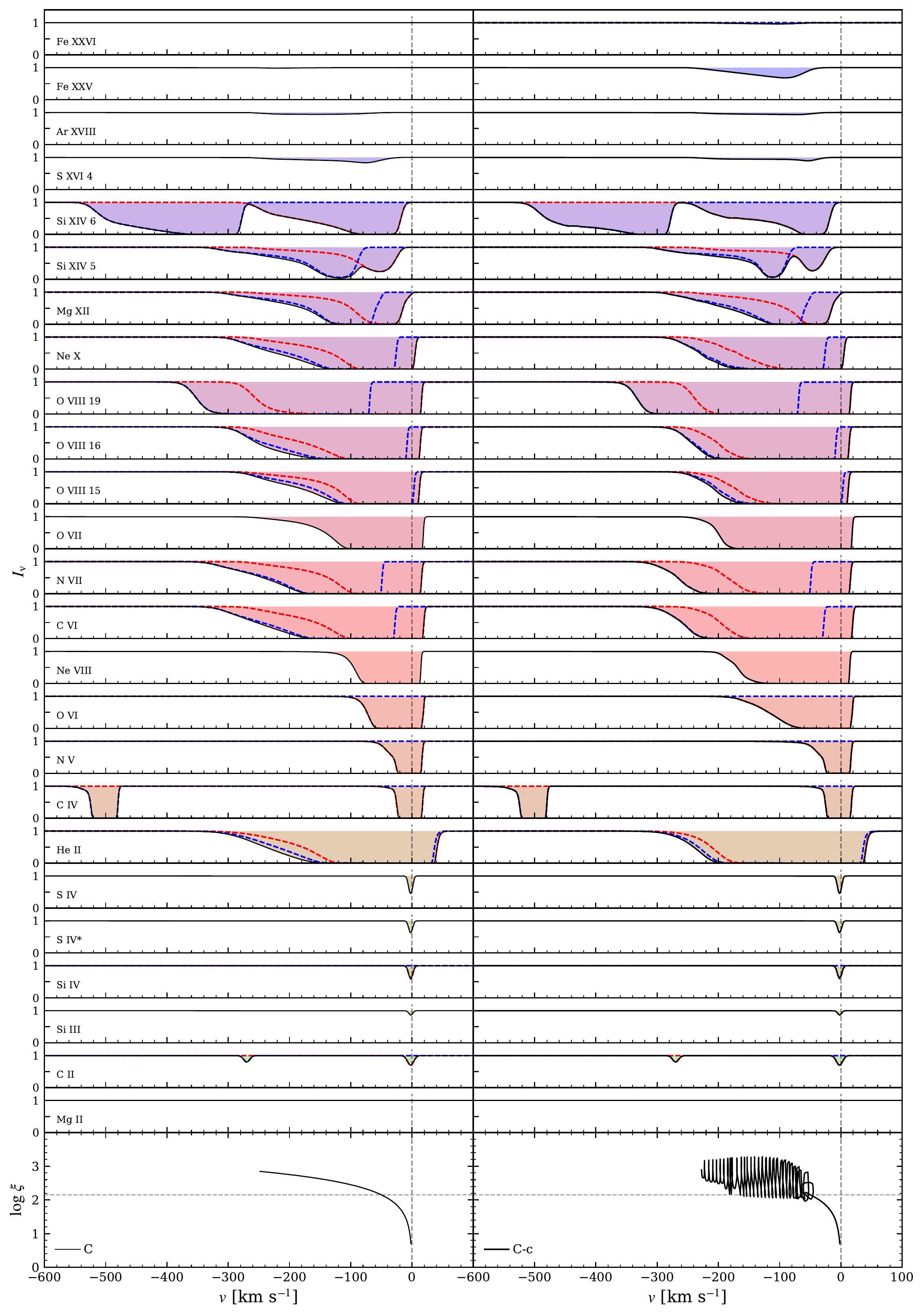}
    \caption{Same as \ref{fig:Bperturb}, but for steady model C compared to clumpy model C-c.}
    \label{fig:Cperturb}
\end{figure*}

In Fig.~\ref{fig:bevuie}, we again show a plot of IE versus $v_B$, this time showing models B-c and C-c over-plotted with models B and C for comparison.
The clumpy cases have a slightly higher overall flow velocity ($v$)
(compare the blue vertical lines; dashed lines are for steady solutions and dotted lines for their clumpy counterparts). 
While higher IE ions show slightly lower $v_B$ for clumpy cases, 
intermediate ions show a very prominent increase in $v_B$ of their absorption 
troughs due to the presence of the clumps. 
This is because these ions (namely, \ion{N}{5}, \ion{O}{6}, and \ion{Ne}{8}) are direct probes of the temperature within the clumps and therefore deepen the absorption at the blue edge.  
The low IE ions, meanwhile, show no change in their $v_B$ values between steady and clumpy models.  Again, these ions form at the base of the flow have negligible opacity within the clumps.

In Figs.~\ref{fig:Bperturb} and \ref{fig:Cperturb}, we compare steady and clumpy models for our entire collection of line profiles. 
As we mentioned above, model C-c is so clumpy that the blue wings of the line profiles remain essentially monotonic.
Rather than produce distinguishable absorption troughs, small intercloud spacings simply result in an increase in the depth of line profiles for most ions. 
There are, however, a few cases of less deep absorption (e.g., \ion{Mg}{12}, \ion{O}{8}, and \ion{Si}{14}), and this occurs also in model B-c as seen most clearly for \ion{Si}{14}~5 in Fig.~\ref{fig:b5}.
To explain this, note that locally, the individual clumps
are isobaric, meaning the clumps are also associated with under-densities corresponding to higher temperature intercloud regions.  For the ions noted, these regions have less opacity compared to smooth solutions where the temperature is intermediate between cloud and intercloud gas temperatures. 
Thus, the column density of ions with peak abundance in intermediate temperature gas can be less in the presence of clumps, and this accounts for the reduced absorption.
Conversely, the higher column of hot gas in the presence of clumps explains the increased absorption in \ion{Fe}{25}, which only forms in high temperature gas.

We conclude here that it is non-trivial to determine the wind structure based on line profiles. Despite the presence of clumps in outflows, line profiles can be smooth and monotonic under certain conditions.
We also note that different velocity components are not necessarily distinctive outflows 
(e.g., flows launched from different geometries with different escape velocities), 
but are instead potentially just different parts of the same flow. 

\section{Discussion} \label{sec:disc}

The conditions for generating outflows that are heated and driven by AGN radiation
place these outflows at parsec scales from the central engine, where the gas is mildly 
bound. Such outflows can be responsible for WAs in AGNs.  In this paper, we
present results from our calculations of synthetic absorption line profiles that are based 
on smooth/clumpy models of thermally driven outflows from AGNs.
Although our calculations are based on 1D radial outflow models,  
we find that each model produces diverse profiles and some profiles are quite complex,
especially for clumpy models.
We classify our line profiles into four major categories: 
1) Gaussian (typical for weak lines, e.g., the \ion{S}{4} line), 
2) boxy (lines with a strong broad core, 
e.g., the \ion{C}{4} line,
3) boxy with an extend blue wing (e.g., \ion{Mg}{12}), and 
4) weak extended (e.g., \ion{Fe}{25}). 
In Figs.~\ref{fig:A21BC}, \ref{fig:Bperturb}, and \ref{fig:Cperturb},
we show examples of how line widths and shapes vary depending on the ionization energy of the absorbing ion. To explore the effects of outflow clumpiness, we 
present the breakdown of a line profile due to \ion{Mg}{12} as an example 
(see Fig.~\ref{fig:breakdown}). This figure shows that clumps can produce 
deeper absorption troughs in the blue wing compared to a smooth flow, 
whereas the cold slow gas at the base of the outflow dominates the line center.

One of the key outflow properties is the terminal velocity. Therefore, we check whether there are
any relations between the maximum outflow velocity in our models and the predicted line properties.
We found that the highly ionized ion species (such as \ion{Fe}{25}) with ionization energies 
above 100~eV trace the fastest part of the outflows. Fig.~\ref{fig:bevsie} shows that line profiles 
due to these very highly ionized species could be strongly blueshifted. Most blueshifts from  these ions 
are comparable to the outflow terminal velocity in our thermally unstable solutions, both steady state and clumpy versions. 
In our thermally stable model A21, however, the blue line 
wing is very extended and weak so that the blue end is difficult to determine. If one then uses the width 
of the strong, often saturated core of the line, the wind velocity can be underestimated by an order of magnitude. 

Our outflow models produce line profiles that are unlike those expected for a spherical flow. 
For example, winds from OB stars can be well approximated as 1D spherical outflows and their absorption 
profiles show maximum absorption  near the highest velocities \citep[e.g.,][]{LC1999}.
Our profiles show maximum absorption near zero velocity which is more characteristic 
of bipolar disk winds in cataclysmic variables \citep[see e.g.][]{Drew1987}. 
We realize that in both winds from OB stars and in winds from cataclysmic variables, line emission 
is important. Nevertheless, this simple comparison illustrates that even smooth spherical winds 
can produce surprising line profiles that at least superficially show similarities to profiles produced 
by axisymmetric disk wind calculations, a good example being the diverse line profiles shown 
by \cite{Giustini2012}. However, in disk winds the diversity is typically due to the line profile dependence on the inclination angle. Here we report 
on the importance of the wind ionization structure \citep[see also the ionization stratification effects on line profiles in][which is our companion paper on clumpy thermally driven disk winds]{Waters2021}.

We note the line profiles of our clumpy solutions are characterized by significantly non-monotonic blue wings 
{\it only} if the separation in velocity space between individual clumps is greater than the thermal width 
of the gas within and between the clumps.  This is the difference between our models B-c and C-c 
(compare Figs.~\ref{fig:Bperturb} and \ref{fig:Cperturb}); the non-monotonic wings in model B-c 
are due to clumps producing blueshifted absorption troughs at velocities outside of the line core.  
Additionally, the small clump spacing in a highly clumped model like C-c, results 
in smaller columns of intercloud gas, which are at temperatures where the opacity of certain lines (\ion{S}{16} 
and \ion{Si}{14}) reach peak values. A close comparison of the left and right panels 
in Figs.~\ref{fig:Cperturb} revealed that these lines are actually weaker in model C-c compared to smooth model~C.  

Future X-ray observatories equipped with micro-calorimeters, such as the X-ray Imaging and Spectroscopy Mission 
\citep[][launch expected in 2022]{2020arXiv200304962X} and the Advanced Telescope for High ENergy Astrophysics \citep[Athena;][launch expected in the 2030s]{2013arXiv1306.2307N} mission should permit distinguishing different 
shapes of absorption line profiles and should also allow characterizing clump-like features in the X-ray spectra 
of nearby AGN such as NGC 5548, if present. Spectra from these observatories will provide a uniform energy resolution 
down to a few eV over a wide energy range, including in the Fe K band, allowing velocity features of only 
a few km s$^{-1}$ to be resolved. Despite the great wealth of current observational data, these future spectral 
resolutions are needed for the precise comparison of our models for WAs due to the relatively small velocities 
involved \citep[e.g., see][for a recent review]{Laha2021}.

Our main conclusion is that thermally driven wind solutions constitute viable models for WAs. 
Therefore, we plan to further develop these 
models, as well as to increase the fidelity of our line profile calculations.
We finish the paper by mentioning just a few of the future developments we have planned.

To better treat line saturation, we plan to implement Lorentzian profiles instead 
of Gaussian profiles. 
In addition, it is of course important to consider the contribution of emission to absorption line profile calculations. 
Indeed, \cite{Lucy1983} showed that non-monotonic velocity profiles in O star winds give rise to saturated 
P~Cygni profiles with flat bottoms. For our future work, we expect that exploration of emission line profiles 
could result in significant emission lines near  $v=0$, which could extend to larger velocities and make absorption appear weaker. 
Studies of  X-ray binaries where AGN-like warm absorbers are sometimes observed 
show that the line emission could be indeed important
\citep[i.e.,][]{Miller2002,Miller2004}.

On the hydrodynamical modeling side, we compared conduction and cooling timescales and found that thermal conduction 
may be important, especially in regions between clumps that have high temperature contrasts 
\cite[the importance of including thermal conduction in local cloud simulations has already been demonstrated; e.g.][]{Proga2015}. Therefore, we plan to assess the effects of thermal conduction on line profiles in our future 
calculations. Finally, the time variability due to clump evolution on observable timescales is insignificant, 
as shown in Fig.~\ref{fig:avg}. This may be indicative of the limitations of our assumption of pure photoionization 
models in thermal equilibrium.  We therefore plan to assess if non-equilibrium effects can be important.

\acknowledgments
Support for this work was provided by the National Aeronautics and Space Administration through Chandra Award Number TM0-21003X issued by the Chandra X-ray Observatory Center, which is operated by the Smithsonian Astrophysical Observatory for and on behalf of the National Aeronautics Space Administration under contract NAS8-03060. This work also was supported by the National Aeronautics Space Administration under ATP grant NNX14AK44G.
MG is supported by the ``Programa de Atracci\'on de Talento'' of the Comunidad de Madrid, grant number 2018-T1/TIC-11733.


\relax 
\providecommand\hyper@newdestlabel[2]{}
\providecommand\HyperFirstAtBeginDocument{\AtBeginDocument}
\HyperFirstAtBeginDocument{\ifx\hyper@anchor\@undefined
\global\let\oldcontentsline\contentsline
\gdef\contentsline#1#2#3#4{\oldcontentsline{#1}{#2}{#3}}
\global\let\oldnewlabel\newlabel
\gdef\newlabel#1#2{\newlabelxx{#1}#2}
\gdef\newlabelxx#1#2#3#4#5#6{\oldnewlabel{#1}{{#2}{#3}}}
\AtEndDocument{\ifx\hyper@anchor\@undefined
\let\contentsline\oldcontentsline
\let\newlabel\oldnewlabel
\fi}
\fi}
\global\let\hyper@last\relax 
\gdef\HyperFirstAtBeginDocument#1{#1}
\providecommand\HyField@AuxAddToFields[1]{}
\providecommand\HyField@AuxAddToCoFields[2]{}
\citation{Reynolds1997,Crenshaw1999}
\citation{Shields1997,Crenshaw1999,Gabel2003,Longinotti2013,Ebrero2016,Fu2017,Mihdipour2017}
\citation{Begelman1983,Woods1996,Waters2021}
\citation{Balsara1993,Doro2008,Doro2008b,Doro2012,Doro2016,Kallman2019}
\citation{Proga2007,Proga2008,Kurosawa2009,Kurosawa2009b}
\citation{Everett2007}
\newlabel{FirstPage}{{}{1}{}{Doc-Start}{}}
\@writefile{toc}{\contentsline {section}{\numberline {1}Introduction}{1}{section.1}\protected@file@percent }
\newlabel{sec:intro}{{1}{1}{Introduction}{section.1}{}}
\citation{Sim2012}
\citation{Kurosawa2009}
\citation{Reynolds1997,Blustin2005,Ricci2017,McKernan2007,Laha2014}
\citation{Sim2012}
\citation{Mizu2019}
\citation{Waters2021}
\citation{Waters2021}
\citation{Field1965}
\citation{Proga2008,Kurosawa2009,Kurosawa2009b,Barai2012,Moscibrodzka2013,Gaspari2013,Takeuchi2013}
\citation{Dannen2020}
\citation{Dannen2020}
\citation{Waters2021}
\citation{Dannen2020}
\citation{Stone2020}
\citation{Dannen2020}
\citation{Mehdipour2015}
\citation{Dannen2019}
\citation{Dannen2020}
\citation{Dannen2020}
\@writefile{toc}{\contentsline {section}{\numberline {2}Models}{2}{section.2}\protected@file@percent }
\newlabel{sec:models}{{2}{2}{Models}{section.2}{}}
\newlabel{eq:Xi}{{1}{2}{Models}{equation.2.1}{}}
\citation{Dannen2020}
\citation{Dannen2020}
\citation{Dannen2020}
\citation{Waters2021}
\citation{Mehdipour2015}
\citation{Balbus1986}
\citation{Balbus1986}
\citation{Dyda2017}
\citation{Dyda2017}
\citation{Kallman2001}
\citation{Dannen2020}
\citation{Dannen2020}
\citation{Waters2021}
\newlabel{tab:steady}{{1}{3}{Introduction}{table.0}{}}
\newlabel{eq:HEP0}{{2}{3}{Models}{equation.2.2}{}}
\newlabel{eq:Ru}{{3}{3}{Models}{equation.2.3}{}}
\citation{Dannen2020}
\newlabel{fig:ps}{{1}{4}{Models}{figure.1}{}}
\newlabel{fig:rad}{{2}{4}{Models}{figure.2}{}}
\citation{Waters2021}
\citation{Dannen2020}
\@writefile{toc}{\contentsline {subsection}{\numberline {2.1}Clumpy models}{5}{subsection.2.1}\protected@file@percent }
\newlabel{sec:clumpy_models}{{2.1}{5}{Clumpy models}{subsection.2.1}{}}
\newlabel{fig:rabund}{{3}{5}{Clumpy models}{figure.3}{}}
\newlabel{eq:bcm5}{{4}{5}{Clumpy models}{equation.2.4}{}}
\newlabel{fig:vabund}{{4}{5}{Clumpy models}{figure.4}{}}
\@writefile{toc}{\contentsline {subsection}{\numberline {2.2}Ionization structure and absorption diagnostics}{5}{subsection.2.2}\protected@file@percent }
\citation{Holczer2007,Behar2009}
\citation{Dyda2017}
\newlabel{eq:kappa}{{5}{6}{Ionization structure and absorption diagnostics}{equation.2.5}{}}
\newlabel{eq:AMD}{{6}{6}{Ionization structure and absorption diagnostics}{equation.2.6}{}}
\newlabel{fig:amd}{{5}{6}{Ionization structure and absorption diagnostics}{figure.5}{}}
\citation{Dannen2020}
\citation{Waters2017}
\gdef \deluxe@table@width@ {700pt}
\citation{Waters2017}
\citation{Blondin1994}
\citation{Waters2017}
\gdef \deluxe@table@width@ {700pt}
\@writefile{toc}{\contentsline {section}{\numberline {3}Methods}{7}{section.3}\protected@file@percent }
\newlabel{sec:method}{{3}{7}{Methods}{section.3}{}}
\newlabel{corrected}{{7}{7}{Methods}{equation.3.7}{}}
\newlabel{eq:tau}{{8}{7}{Methods}{equation.3.8}{}}
\newlabel{eq:kappanu}{{9}{7}{Methods}{equation.3.9}{}}
\newlabel{eq:phi}{{10}{7}{Methods}{equation.3.10}{}}
\newlabel{eq:inu}{{11}{7}{Methods}{equation.3.11}{}}
\newlabel{tab:all}{{2}{8}{Methods}{table.0}{}}
\@writefile{toc}{\contentsline {section}{\numberline {4}Results}{8}{section.4}\protected@file@percent }
\newlabel{sec:res}{{4}{8}{Results}{section.4}{}}
\@writefile{toc}{\contentsline {subsection}{\numberline {4.1}Steady state wind solutions}{8}{subsection.4.1}\protected@file@percent }
\newlabel{sec:res_steady}{{4.1}{8}{Steady state wind solutions}{subsection.4.1}{}}
\newlabel{tab:nion}{{3}{9}{Methods}{table.0}{}}
\newlabel{fig:Nion_vs_vB}{{6}{9}{Methods}{figure.6}{}}
\newlabel{fig:all5}{{7}{10}{Steady state wind solutions}{figure.7}{}}
\newlabel{fig:A21BC}{{8}{11}{Steady state wind solutions}{figure.8}{}}
\newlabel{fig:bevsie}{{9}{12}{Steady state wind solutions}{figure.9}{}}
\@writefile{toc}{\contentsline {subsection}{\numberline {4.2}Clumpy wind solutions}{12}{subsection.4.2}\protected@file@percent }
\newlabel{sec:res_clumpy}{{4.2}{12}{Clumpy wind solutions}{subsection.4.2}{}}
\citation{Arav1997}
\newlabel{fig:b5}{{10}{13}{Steady state wind solutions}{figure.10}{}}
\newlabel{fig:breakdown}{{11}{13}{Steady state wind solutions}{figure.11}{}}
\newlabel{fig:avg}{{12}{14}{Steady state wind solutions}{figure.12}{}}
\newlabel{fig:bevuie}{{13}{14}{Steady state wind solutions}{figure.13}{}}
\newlabel{tcool}{{12}{14}{Clumpy wind solutions}{equation.4.12}{}}
\newlabel{fig:Bperturb}{{14}{15}{Clumpy wind solutions}{figure.14}{}}
\newlabel{fig:Cperturb}{{15}{16}{Clumpy wind solutions}{figure.15}{}}
\citation{LC1999}
\citation{Drew1987}
\citation{Giustini2012}
\citation{Waters2021}
\@writefile{toc}{\contentsline {section}{\numberline {5}Discussion}{17}{section.5}\protected@file@percent }
\newlabel{sec:disc}{{5}{17}{Discussion}{section.5}{}}
\citation{2020arXiv200304962X}
\citation{2013arXiv1306.2307N}
\citation{Laha2021}
\citation{Lucy1983}
\citation{Miller2002,Miller2004}
\citation{Proga2015}
\bibcite{Arav1997}{{1}{1997}{{{Arav} {et~al.}}}{{{Arav}, {Barlow}, {Laor}, \& {Blandford}}}}
\bibcite{Balbus1986}{{2}{1986}{{Balbus}}{{}}}
\bibcite{Balsara1993}{{3}{1993}{{{Balsara} \& {Krolik}}}{{}}}
\bibcite{Barai2012}{{4}{2012}{{{Barai} {et~al.}}}{{{Barai}, {Proga}, \& {Nagamine}}}}
\bibcite{Begelman1983}{{5}{1983}{{{Begelman} {et~al.}}}{{{Begelman}, {McKee}, \& {Shields}}}}
\bibcite{Behar2009}{{6}{2009}{{Behar}}{{}}}
\bibcite{Blondin1994}{{7}{1994}{{Blondin}}{{}}}
\bibcite{Blustin2005}{{8}{2005}{{{Blustin} {et~al.}}}{{{Blustin}, {Page}, {Fuerst}, {Brand uardi-Raymont}, \& {Ashton}}}}
\bibcite{Crenshaw1999}{{9}{1999}{{{Crenshaw} {et~al.}}}{{{Crenshaw}, {Kraemer}, {Boggess}, {Maran}, {Mushotzky}, \& {Wu}}}}
\bibcite{Dannen2019}{{10}{2019}{{{Dannen} {et~al.}}}{{{Dannen}, {Proga}, {Kallman}, \& {Waters}}}}
\bibcite{Dannen2020}{{11}{2020}{{{Dannen} {et~al.}}}{{{Dannen}, {Proga}, {Waters}, \& {Dyda}}}}
\bibcite{Doro2012}{{12}{2012}{{{Dorodnitsyn} {et~al.}}}{{{Dorodnitsyn}, {Kallman}, \& {Bisnovatyi-Kogan}}}}
\bibcite{Doro2008}{{13}{2008{a}}{{{Dorodnitsyn} {et~al.}}}{{{Dorodnitsyn}, {Kallman}, \& {Proga}}}}
\bibcite{Doro2008b}{{14}{2008{b}}{{{Dorodnitsyn} {et~al.}}}{{{Dorodnitsyn}, {Kallman}, \& {Proga}}}}
\bibcite{Doro2016}{{15}{2016}{{{Dorodnitsyn} {et~al.}}}{{{Dorodnitsyn}, {Kallman}, \& {Proga}}}}
\bibcite{Drew1987}{{16}{1987}{{Drew}}{{}}}
\bibcite{Dyda2017}{{17}{2017}{{{Dyda} {et~al.}}}{{{Dyda}, {Dannen}, {Waters}, \& {Proga}}}}
\bibcite{Ebrero2016}{{18}{2016}{{{Ebrero} {et~al.}}}{{{Ebrero}, {Kriss}, {Kaastra}, \& {Ely}}}}
\bibcite{Everett2007}{{19}{2007}{{{Everett} \& {Murray}}}{{}}}
\bibcite{Field1965}{{20}{1965}{{Field}}{{}}}
\bibcite{Fu2017}{{21}{2017}{{{Fu} {et~al.}}}{{{Fu}, {Zhang}, {Sun}, {Niu}, \& {Ji}}}}
\bibcite{Gabel2003}{{22}{2003}{{{Gabel} {et~al.}}}{{{Gabel}, {Crenshaw}, {Kraemer}, {Brandt}, {George}, {Hamann}, {Kaiser}, {Kaspi}, {Kriss}, {Mathur}, {Mushotzky}, {Nandra}, {Netzer}, {Peterson}, {Shields}, {Turner}, \& {Zheng}}}}
\bibcite{Gaspari2013}{{23}{2013}{{{Gaspari} {et~al.}}}{{{Gaspari}, {Ruszkowski}, \& {Oh}}}}
\bibcite{Giustini2012}{{24}{2012}{{{Giustini} \& {Proga}}}{{}}}
\bibcite{Holczer2007}{{25}{2007}{{Holczer {et~al.}}}{{Holczer, Behar, \& Kaspi}}}
\bibcite{Kallman2001}{{26}{2001}{{{Kallman} \& {Bautista}}}{{}}}
\bibcite{Kallman2019}{{27}{2019}{{{Kallman} \& {Dorodnitsyn}}}{{}}}
\bibcite{Kurosawa2009}{{28}{2009{a}}{{{Kurosawa} \& {Proga}}}{{}}}
\bibcite{Kurosawa2009b}{{29}{2009{b}}{{{Kurosawa} \& {Proga}}}{{}}}
\bibcite{Laha2014}{{30}{2014}{{{Laha} {et~al.}}}{{{Laha}, {Guainazzi}, {Dewangan}, {Chakravorty}, \& {Kembhavi}}}}
\bibcite{LC1999}{{31}{1999}{{{Lamers} \& {Cassinelli}}}{{}}}
\bibcite{Longinotti2013}{{32}{2013}{{{Longinotti} {et~al.}}}{{{Longinotti}, {Krongold}, {Kriss}, {Ely}, {Gallo}, {Grupe}, {Komossa}, {Mathur}, \& {Pradhan}}}}
\bibcite{Lucy1983}{{33}{1983}{{Lucy}}{{}}}
\bibcite{McKernan2007}{{34}{2007}{{{McKernan} {et~al.}}}{{{McKernan}, {Yaqoob}, \& {Reynolds}}}}
\bibcite{Mehdipour2015}{{35}{2015}{{{Mehdipour} {et~al.}}}{{{Mehdipour}, {Kaastra}, {Kriss}, {Cappi}, {Petrucci}, {Steenbrugge}, {Arav}, {Behar}, {Bianchi}, {Boissay}, {Brand uardi-Raymont}, {Costantini}, {Ebrero}, {Di Gesu}, {Harrison}, {Kaspi}, {De Marco}, {Matt}, {Paltani}, {Peterson}, {Ponti}, {Pozo Nu{\~n}ez}, {De Rosa}, {Ursini}, {de Vries}, {Walton}, \& {Whewell}}}}
\bibcite{Mihdipour2017}{{36}{2017}{{{Mehdipour} {et~al.}}}{{{Mehdipour}, {Kaastra}, {Kriss}, {Arav}, {Behar}, {Bianchi}, {Branduardi-Raymont}, {Cappi}, {Costantini}, {Ebrero}, {Di Gesu}, {Kaspi}, {Mao}, {De Marco}, {Matt}, {Paltani}, {Peretz}, {Peterson}, {Petrucci}, {Pinto}, {Ponti}, {Ursini}, {de Vries}, \& {Walton}}}}
\bibcite{Miller2002}{{37}{2002}{{{Miller} {et~al.}}}{{{Miller}, {Fabian}, {Wijnands}, {Remillard}, {Wojdowski}, {Schulz}, {Di Matteo}, {Marshall}, {Canizares}, {Pooley}, \& {Lewin}}}}
\bibcite{Miller2004}{{38}{2004}{{{Miller} {et~al.}}}{{{Miller}, {Raymond}, {Fabian}, {Homan}, {Nowak}, {Wijnands}, {van der Klis}, {Belloni}, {Tomsick}, {Smith}, {Charles}, \& {Lewin}}}}
\bibcite{Mizu2019}{{39}{2019}{{{Mizumoto} {et~al.}}}{{{Mizumoto}, {Done}, {Tomaru}, \& {Edwards}}}}
\bibcite{Moscibrodzka2013}{{40}{2013}{{{Mo{\'s}cibrodzka} \& {Proga}}}{{}}}
\bibcite{2013arXiv1306.2307N}{{41}{2013}{{{Nandra} {et~al.}}}{{{Nandra}, {Barret}, {Barcons}, {Fabian}, {den Herder}, {Piro}, {Watson}, {Adami}, {Aird}, {Afonso}, {Alexander}, {Argiroffi}, {Amati}, {Arnaud}, {Atteia}, {Audard}, {Badenes}, {Ballet}, {Ballo}, {Bamba}, {Bhardwaj}, {Stefano Battistelli}, {Becker}, {De Becker}, {Behar}, {Bianchi}, {Biffi}, {B{\^\i }rzan}, {Bocchino}, {Bogdanov}, {Boirin}, {Boller}, {Borgani}, {Borm}, {Bouch{\'e}}, {Bourdin}, {Bower}, {Braito}, {Branchini}, {Branduardi-Raymont}, {Bregman}, {Brenneman}, {Brightman}, {Br{\"u}ggen}, {Buchner}, {Bulbul}, {Brusa}, {Bursa}, {Caccianiga}, {Cackett}, {Campana}, {Cappelluti}, {Cappi}, {Carrera}, {Ceballos}, {Christensen}, {Chu}, {Churazov}, {Clerc}, {Corbel}, {Corral}, {Comastri}, {Costantini}, {Croston}, {Dadina}, {D'Ai}, {Decourchelle}, {Della Ceca}, {Dennerl}, {Dolag}, {Done}, {Dovciak}, {Drake}, {Eckert}, {Edge}, {Ettori}, {Ezoe}, {Feigelson}, {Fender}, {Feruglio}, {Finoguenov}, {Fiore}, {Galeazzi}, {Gallagher}, {Gandhi}, {Gaspari}, {Gastaldello}, {Georgakakis}, {Georgantopoulos}, {Gilfanov}, {Gitti}, {Gladstone}, {Goosmann}, {Gosset}, {Grosso}, {Guedel}, {Guerrero}, {Haberl}, {Hardcastle}, {Heinz}, {Alonso Herrero}, {Herv{\'e}}, {Holmstrom}, {Iwasawa}, {Jonker}, {Kaastra}, {Kara}, {Karas}, {Kastner}, {King}, {Kosenko}, {Koutroumpa}, {Kraft}, {Kreykenbohm}, {Lallement}, {Lanzuisi}, {Lee}, {Lemoine-Goumard}, {Lobban}, {Lodato}, {Lovisari}, {Lotti}, {McCharthy}, {McNamara}, {Maggio}, {Maiolino}, {De Marco}, {de Martino}, {Mateos}, {Matt}, {Maughan}, {Mazzotta}, {Mendez}, {Merloni}, {Micela}, {Miceli}, {Mignani}, {Miller}, {Miniutti}, {Molendi}, {Montez}, {Moretti}, {Motch}, {Naz{\'e}}, {Nevalainen}, {Nicastro}, {Nulsen}, {Ohashi}, {O'Brien}, {Osborne}, {Oskinova}, {Pacaud}, {Paerels}, {Page}, {Papadakis}, {Pareschi}, {Petre}, {Petrucci}, {Piconcelli}, {Pillitteri}, {Pinto}, {de Plaa}, {Pointecouteau}, {Ponman}, {Ponti}, {Porquet}, {Pounds}, {Pratt}, {Predehl}, {Proga}, {Psaltis}, {Rafferty}, {Ramos-Ceja}, {Ranalli}, {Rasia}, {Rau}, {Rauw}, {Rea}, {Read}, {Reeves}, {Reiprich}, {Renaud}, {Reynolds}, {Risaliti}, {Rodriguez}, {Rodriguez Hidalgo}, {Roncarelli}, {Rosario}, {Rossetti}, {Rozanska}, {Rovilos}, {Salvaterra}, {Salvato}, {Di Salvo}, {Sanders}, {Sanz-Forcada}, {Schawinski}, {Schaye}, {Schwope}, {Sciortino}, {Severgnini}, {Shankar}, {Sijacki}, {Sim}, {Schmid}, {Smith}, {Steiner}, {Stelzer}, {Stewart}, {Strohmayer}, {Str{\"u}der}, {Sun}, {Takei}, {Tatischeff}, {Tiengo}, {Tombesi}, {Trinchieri}, {Tsuru}, {Ud-Doula}, {Ursino}, {Valencic}, {Vanzella}, {Vaughan}, {Vignali}, {Vink}, {Vito}, {Volonteri}, {Wang}, {Webb}, {Willingale}, {Wilms}, {Wise}, {Worrall}, {Young}, {Zampieri}, {In't Zand}, {Zane}, {Zezas}, {Zhang}, \& {Zhuravleva}}}}
\bibcite{Proga2007}{{42}{2007}{{Proga}}{{}}}
\bibcite{Proga2008}{{43}{2008}{{{Proga} {et~al.}}}{{{Proga}, {Ostriker}, \& {Kurosawa}}}}
\bibcite{Proga2015}{{44}{2015}{{{Proga} \& {Waters}}}{{}}}
\bibcite{Reynolds1997}{{45}{1997}{{Reynolds}}{{}}}
\bibcite{Ricci2017}{{46}{2017}{{{Ricci} {et~al.}}}{{{Ricci}, {Trakhtenbrot}, {Koss}, {Ueda}, {Schawinski}, {Oh}, {Lamperti}, {Mushotzky}, {Treister}, {Ho}, {Weigel}, {Bauer}, {Paltani}, {Fabian}, {Xie}, \& {Gehrels}}}}
\bibcite{Shields1997}{{47}{1997}{{{Shields} \& {Hamann}}}{{}}}
\bibcite{Sim2012}{{48}{2012}{{{Sim} {et~al.}}}{{{Sim}, {Proga}, {Kurosawa}, {Long}, {Miller}, \& {Turner}}}}
\bibcite{Stone2020}{{49}{2020}{{{Stone} {et~al.}}}{{{Stone}, {Tomida}, {White}, \& {Felker}}}}
\bibcite{Takeuchi2013}{{50}{2013}{{{Takeuchi} {et~al.}}}{{{Takeuchi}, {Ohsuga}, \& {Mineshige}}}}
\bibcite{Waters2021}{{51}{2021}{{{Waters} {et~al.}}}{{{Waters}, {Proga}, \& {Dannen}}}}
\bibcite{Waters2017}{{52}{2017}{{{Waters} {et~al.}}}{{{Waters}, {Proga}, {Dannen}, \& {Kallman}}}}
\bibcite{Woods1996}{{53}{1996}{{{Woods} {et~al.}}}{{{Woods}, {Klein}, {Castor}, {McKee}, \& {Bell}}}}
\bibcite{2020arXiv200304962X}{{54}{2020}{{XRISM Science Team}}{{}}}
\bibstyle{aasjournal}
\newlabel{LastPage}{{}{20}{}{}{}}


\begin{thebibliography}{}
\expandafter\ifx\csname natexlab\endcsname\relax\def\natexlab#1{#1}\fi
\providecommand{\url}[1]{\href{#1}{#1}}
\providecommand{\dodoi}[1]{doi:~\href{http://doi.org/#1}{\nolinkurl{#1}}}
\providecommand{\doeprint}[1]{\href{http://ascl.net/#1}{\nolinkurl{http://ascl.net/#1}}}
\providecommand{\doarXiv}[1]{\href{https://arxiv.org/abs/#1}{\nolinkurl{https://arxiv.org/abs/#1}}}

\bibitem[{{Arav} {et~al.}(1997){Arav}, {Barlow}, {Laor}, \&
  {Blandford}}]{Arav1997}
{Arav}, N., {Barlow}, T.~A., {Laor}, A., \& {Blandford}, R.~D. 1997, \mnras,
  288, 1015, \dodoi{10.1093/mnras/288.4.1015}

\bibitem[{{Balbus}(1986)}]{Balbus1986}
{Balbus}, S.~A. 1986, \apjl, 303, L79, \dodoi{10.1086/184657}

\bibitem[{{Balsara} \& {Krolik}(1993)}]{Balsara1993}
{Balsara}, D.~S., \& {Krolik}, J.~H. 1993, \apj, 402, 109,
  \dodoi{10.1086/172116}

\bibitem[{{Barai} {et~al.}(2012){Barai}, {Proga}, \& {Nagamine}}]{Barai2012}
{Barai}, P., {Proga}, D., \& {Nagamine}, K. 2012, \mnras, 424, 728,
  \dodoi{10.1111/j.1365-2966.2012.21260.x}

\bibitem[{{Begelman} {et~al.}(1983){Begelman}, {McKee}, \&
  {Shields}}]{Begelman1983}
{Begelman}, M.~C., {McKee}, C.~F., \& {Shields}, G.~A. 1983, \apj, 271, 70,
  \dodoi{10.1086/161178}

\bibitem[{Behar(2009)}]{Behar2009}
Behar, E. 2009, The Astrophysical Journal, 703, 1346,
  \dodoi{10.1088/0004-637x/703/2/1346}

\bibitem[{{Blondin}(1994)}]{Blondin1994}
{Blondin}, J.~M. 1994, \apj, 435, 756, \dodoi{10.1086/174853}

\bibitem[{{Blustin} {et~al.}(2005){Blustin}, {Page}, {Fuerst}, {Brand
  uardi-Raymont}, \& {Ashton}}]{Blustin2005}
{Blustin}, A.~J., {Page}, M.~J., {Fuerst}, S.~V., {Brand uardi-Raymont}, G., \&
  {Ashton}, C.~E. 2005, \aap, 431, 111, \dodoi{10.1051/0004-6361:20041775}

\bibitem[{{Crenshaw} {et~al.}(1999){Crenshaw}, {Kraemer}, {Boggess}, {Maran},
  {Mushotzky}, \& {Wu}}]{Crenshaw1999}
{Crenshaw}, D.~M., {Kraemer}, S.~B., {Boggess}, A., {et~al.} 1999, \apj, 516,
  750, \dodoi{10.1086/307144}

\bibitem[{{Dannen} {et~al.}(2019){Dannen}, {Proga}, {Kallman}, \&
  {Waters}}]{Dannen2019}
{Dannen}, R.~C., {Proga}, D., {Kallman}, T.~R., \& {Waters}, T. 2019, \apj,
  882, 99, \dodoi{10.3847/1538-4357/ab340b}

\bibitem[{{Dannen} {et~al.}(2020){Dannen}, {Proga}, {Waters}, \&
  {Dyda}}]{Dannen2020}
{Dannen}, R.~C., {Proga}, D., {Waters}, T., \& {Dyda}, S. 2020, \apjl, 893,
  L34, \dodoi{10.3847/2041-8213/ab87a5}

\bibitem[{{Dorodnitsyn} {et~al.}(2012){Dorodnitsyn}, {Kallman}, \&
  {Bisnovatyi-Kogan}}]{Doro2012}
{Dorodnitsyn}, A., {Kallman}, T., \& {Bisnovatyi-Kogan}, G.~S. 2012, \apj, 747,
  8, \dodoi{10.1088/0004-637X/747/1/8}

\bibitem[{{Dorodnitsyn} {et~al.}(2008{\natexlab{a}}){Dorodnitsyn}, {Kallman},
  \& {Proga}}]{Doro2008}
{Dorodnitsyn}, A., {Kallman}, T., \& {Proga}, D. 2008{\natexlab{a}}, \apjl,
  675, L5, \dodoi{10.1086/529374}

\bibitem[{{Dorodnitsyn} {et~al.}(2008{\natexlab{b}}){Dorodnitsyn}, {Kallman},
  \& {Proga}}]{Doro2008b}
---. 2008{\natexlab{b}}, \apj, 687, 97, \dodoi{10.1086/591418}

\bibitem[{{Dorodnitsyn} {et~al.}(2016){Dorodnitsyn}, {Kallman}, \&
  {Proga}}]{Doro2016}
---. 2016, \apj, 819, 115, \dodoi{10.3847/0004-637X/819/2/115}

\bibitem[{{Drew}(1987)}]{Drew1987}
{Drew}, J.~E. 1987, \mnras, 224, 595, \dodoi{10.1093/mnras/224.3.595}

\bibitem[{{Dyda} {et~al.}(2017){Dyda}, {Dannen}, {Waters}, \&
  {Proga}}]{Dyda2017}
{Dyda}, S., {Dannen}, R., {Waters}, T., \& {Proga}, D. 2017, \mnras, 467, 4161,
  \dodoi{10.1093/mnras/stx406}

\bibitem[{{Ebrero} {et~al.}(2016){Ebrero}, {Kriss}, {Kaastra}, \&
  {Ely}}]{Ebrero2016}
{Ebrero}, J., {Kriss}, G.~A., {Kaastra}, J.~S., \& {Ely}, J.~C. 2016, \aap,
  586, A72, \dodoi{10.1051/0004-6361/201527495}

\bibitem[{{Everett} \& {Murray}(2007)}]{Everett2007}
{Everett}, J.~E., \& {Murray}, N. 2007, \apj, 656, 93, \dodoi{10.1086/510324}

\bibitem[{{Field}(1965)}]{Field1965}
{Field}, G.~B. 1965, \apj, 142, 531, \dodoi{10.1086/148317}

\bibitem[{{Fu} {et~al.}(2017){Fu}, {Zhang}, {Sun}, {Niu}, \& {Ji}}]{Fu2017}
{Fu}, X.-D., {Zhang}, S.-N., {Sun}, W., {Niu}, S., \& {Ji}, L. 2017, Research
  in Astronomy and Astrophysics, 17, 095, \dodoi{10.1088/1674-4527/17/9/95}

\bibitem[{{Gabel} {et~al.}(2003){Gabel}, {Crenshaw}, {Kraemer}, {Brandt},
  {George}, {Hamann}, {Kaiser}, {Kaspi}, {Kriss}, {Mathur}, {Mushotzky},
  {Nandra}, {Netzer}, {Peterson}, {Shields}, {Turner}, \& {Zheng}}]{Gabel2003}
{Gabel}, J.~R., {Crenshaw}, D.~M., {Kraemer}, S.~B., {et~al.} 2003, \apj, 583,
  178, \dodoi{10.1086/345096}

\bibitem[{{Gaspari} {et~al.}(2013){Gaspari}, {Ruszkowski}, \&
  {Oh}}]{Gaspari2013}
{Gaspari}, M., {Ruszkowski}, M., \& {Oh}, S.~P. 2013, \mnras, 432, 3401,
  \dodoi{10.1093/mnras/stt692}

\bibitem[{{Giustini} \& {Proga}(2012)}]{Giustini2012}
{Giustini}, M., \& {Proga}, D. 2012, \apj, 758, 70,
  \dodoi{10.1088/0004-637X/758/1/70}

\bibitem[{Holczer {et~al.}(2007)Holczer, Behar, \& Kaspi}]{Holczer2007}
Holczer, T., Behar, E., \& Kaspi, S. 2007, The Astrophysical Journal, 663, 799,
  \dodoi{10.1086/518416}

\bibitem[{{Kallman} \& {Bautista}(2001)}]{Kallman2001}
{Kallman}, T., \& {Bautista}, M. 2001, \apjs, 133, 221, \dodoi{10.1086/319184}

\bibitem[{{Kallman} \& {Dorodnitsyn}(2019)}]{Kallman2019}
{Kallman}, T., \& {Dorodnitsyn}, A. 2019, \apj, 884, 111,
  \dodoi{10.3847/1538-4357/ab40aa}

\bibitem[{{Kurosawa} \& {Proga}(2009{\natexlab{a}})}]{Kurosawa2009}
{Kurosawa}, R., \& {Proga}, D. 2009{\natexlab{a}}, \apj, 693, 1929,
  \dodoi{10.1088/0004-637X/693/2/1929}

\bibitem[{{Kurosawa} \& {Proga}(2009{\natexlab{b}})}]{Kurosawa2009b}
---. 2009{\natexlab{b}}, \mnras, 397, 1791,
  \dodoi{10.1111/j.1365-2966.2009.15084.x}

\bibitem[{{Laha} {et~al.}(2014){Laha}, {Guainazzi}, {Dewangan}, {Chakravorty},
  \& {Kembhavi}}]{Laha2014}
{Laha}, S., {Guainazzi}, M., {Dewangan}, G.~C., {Chakravorty}, S., \&
  {Kembhavi}, A.~K. 2014, \mnras, 441, 2613, \dodoi{10.1093/mnras/stu669}

\bibitem[{{Lamers} \& {Cassinelli}(1999)}]{LC1999}
{Lamers}, H. J.~G.~L.~M., \& {Cassinelli}, J.~P. 1999, {Introduction to Stellar
  Winds} (Cambridge)

\bibitem[{{Longinotti} {et~al.}(2013){Longinotti}, {Krongold}, {Kriss}, {Ely},
  {Gallo}, {Grupe}, {Komossa}, {Mathur}, \& {Pradhan}}]{Longinotti2013}
{Longinotti}, A.~L., {Krongold}, Y., {Kriss}, G.~A., {et~al.} 2013, \apj, 766,
  104, \dodoi{10.1088/0004-637X/766/2/104}

\bibitem[{{Lucy}(1983)}]{Lucy1983}
{Lucy}, L.~B. 1983, \apj, 274, 372, \dodoi{10.1086/161453}

\bibitem[{{McKernan} {et~al.}(2007){McKernan}, {Yaqoob}, \&
  {Reynolds}}]{McKernan2007}
{McKernan}, B., {Yaqoob}, T., \& {Reynolds}, C.~S. 2007, \mnras, 379, 1359,
  \dodoi{10.1111/j.1365-2966.2007.11993.x}

\bibitem[{{Mehdipour} {et~al.}(2015){Mehdipour}, {Kaastra}, {Kriss}, {Cappi},
  {Petrucci}, {Steenbrugge}, {Arav}, {Behar}, {Bianchi}, {Boissay}, {Brand
  uardi-Raymont}, {Costantini}, {Ebrero}, {Di Gesu}, {Harrison}, {Kaspi}, {De
  Marco}, {Matt}, {Paltani}, {Peterson}, {Ponti}, {Pozo Nu{\~n}ez}, {De Rosa},
  {Ursini}, {de Vries}, {Walton}, \& {Whewell}}]{Mehdipour2015}
{Mehdipour}, M., {Kaastra}, J.~S., {Kriss}, G.~A., {et~al.} 2015, \aap, 575,
  A22, \dodoi{10.1051/0004-6361/201425373}

\bibitem[{{Mehdipour} {et~al.}(2017){Mehdipour}, {Kaastra}, {Kriss}, {Arav},
  {Behar}, {Bianchi}, {Branduardi-Raymont}, {Cappi}, {Costantini}, {Ebrero},
  {Di Gesu}, {Kaspi}, {Mao}, {De Marco}, {Matt}, {Paltani}, {Peretz},
  {Peterson}, {Petrucci}, {Pinto}, {Ponti}, {Ursini}, {de Vries}, \&
  {Walton}}]{Mihdipour2017}
---. 2017, \aap, 607, A28, \dodoi{10.1051/0004-6361/201731175}

\bibitem[{{Miller} {et~al.}(2002){Miller}, {Fabian}, {Wijnands}, {Remillard},
  {Wojdowski}, {Schulz}, {Di Matteo}, {Marshall}, {Canizares}, {Pooley}, \&
  {Lewin}}]{Miller2002}
{Miller}, J.~M., {Fabian}, A.~C., {Wijnands}, R., {et~al.} 2002, \apj, 578,
  348, \dodoi{10.1086/342466}

\bibitem[{{Miller} {et~al.}(2004){Miller}, {Raymond}, {Fabian}, {Homan},
  {Nowak}, {Wijnands}, {van der Klis}, {Belloni}, {Tomsick}, {Smith},
  {Charles}, \& {Lewin}}]{Miller2004}
{Miller}, J.~M., {Raymond}, J., {Fabian}, A.~C., {et~al.} 2004, \apj, 601, 450,
  \dodoi{10.1086/380196}

\bibitem[{{Mizumoto} {et~al.}(2019){Mizumoto}, {Done}, {Tomaru}, \&
  {Edwards}}]{Mizu2019}
{Mizumoto}, M., {Done}, C., {Tomaru}, R., \& {Edwards}, I. 2019, \mnras, 489,
  1152, \dodoi{10.1093/mnras/stz2225}

\bibitem[{{Mo{\'s}cibrodzka} \& {Proga}(2013)}]{Moscibrodzka2013}
{Mo{\'s}cibrodzka}, M., \& {Proga}, D. 2013, \apj, 767, 156,
  \dodoi{10.1088/0004-637X/767/2/156}

\bibitem[{{Nandra} {et~al.}(2013){Nandra}, {Barret}, {Barcons}, {Fabian}, {den
  Herder}, {Piro}, {Watson}, {Adami}, {Aird}, {Afonso}, {Alexander},
  {Argiroffi}, {Amati}, {Arnaud}, {Atteia}, {Audard}, {Badenes}, {Ballet},
  {Ballo}, {Bamba}, {Bhardwaj}, {Stefano Battistelli}, {Becker}, {De Becker},
  {Behar}, {Bianchi}, {Biffi}, {B{\^\i}rzan}, {Bocchino}, {Bogdanov}, {Boirin},
  {Boller}, {Borgani}, {Borm}, {Bouch{\'e}}, {Bourdin}, {Bower}, {Braito},
  {Branchini}, {Branduardi-Raymont}, {Bregman}, {Brenneman}, {Brightman},
  {Br{\"u}ggen}, {Buchner}, {Bulbul}, {Brusa}, {Bursa}, {Caccianiga},
  {Cackett}, {Campana}, {Cappelluti}, {Cappi}, {Carrera}, {Ceballos},
  {Christensen}, {Chu}, {Churazov}, {Clerc}, {Corbel}, {Corral}, {Comastri},
  {Costantini}, {Croston}, {Dadina}, {D'Ai}, {Decourchelle}, {Della Ceca},
  {Dennerl}, {Dolag}, {Done}, {Dovciak}, {Drake}, {Eckert}, {Edge}, {Ettori},
  {Ezoe}, {Feigelson}, {Fender}, {Feruglio}, {Finoguenov}, {Fiore}, {Galeazzi},
  {Gallagher}, {Gandhi}, {Gaspari}, {Gastaldello}, {Georgakakis},
  {Georgantopoulos}, {Gilfanov}, {Gitti}, {Gladstone}, {Goosmann}, {Gosset},
  {Grosso}, {Guedel}, {Guerrero}, {Haberl}, {Hardcastle}, {Heinz}, {Alonso
  Herrero}, {Herv{\'e}}, {Holmstrom}, {Iwasawa}, {Jonker}, {Kaastra}, {Kara},
  {Karas}, {Kastner}, {King}, {Kosenko}, {Koutroumpa}, {Kraft}, {Kreykenbohm},
  {Lallement}, {Lanzuisi}, {Lee}, {Lemoine-Goumard}, {Lobban}, {Lodato},
  {Lovisari}, {Lotti}, {McCharthy}, {McNamara}, {Maggio}, {Maiolino}, {De
  Marco}, {de Martino}, {Mateos}, {Matt}, {Maughan}, {Mazzotta}, {Mendez},
  {Merloni}, {Micela}, {Miceli}, {Mignani}, {Miller}, {Miniutti}, {Molendi},
  {Montez}, {Moretti}, {Motch}, {Naz{\'e}}, {Nevalainen}, {Nicastro}, {Nulsen},
  {Ohashi}, {O'Brien}, {Osborne}, {Oskinova}, {Pacaud}, {Paerels}, {Page},
  {Papadakis}, {Pareschi}, {Petre}, {Petrucci}, {Piconcelli}, {Pillitteri},
  {Pinto}, {de Plaa}, {Pointecouteau}, {Ponman}, {Ponti}, {Porquet}, {Pounds},
  {Pratt}, {Predehl}, {Proga}, {Psaltis}, {Rafferty}, {Ramos-Ceja}, {Ranalli},
  {Rasia}, {Rau}, {Rauw}, {Rea}, {Read}, {Reeves}, {Reiprich}, {Renaud},
  {Reynolds}, {Risaliti}, {Rodriguez}, {Rodriguez Hidalgo}, {Roncarelli},
  {Rosario}, {Rossetti}, {Rozanska}, {Rovilos}, {Salvaterra}, {Salvato}, {Di
  Salvo}, {Sanders}, {Sanz-Forcada}, {Schawinski}, {Schaye}, {Schwope},
  {Sciortino}, {Severgnini}, {Shankar}, {Sijacki}, {Sim}, {Schmid}, {Smith},
  {Steiner}, {Stelzer}, {Stewart}, {Strohmayer}, {Str{\"u}der}, {Sun}, {Takei},
  {Tatischeff}, {Tiengo}, {Tombesi}, {Trinchieri}, {Tsuru}, {Ud-Doula},
  {Ursino}, {Valencic}, {Vanzella}, {Vaughan}, {Vignali}, {Vink}, {Vito},
  {Volonteri}, {Wang}, {Webb}, {Willingale}, {Wilms}, {Wise}, {Worrall},
  {Young}, {Zampieri}, {In't Zand}, {Zane}, {Zezas}, {Zhang}, \&
  {Zhuravleva}}]{2013arXiv1306.2307N}
{Nandra}, K., {Barret}, D., {Barcons}, X., {et~al.} 2013, arXiv e-prints,
  arXiv:1306.2307.
\newblock \doarXiv{1306.2307}

\bibitem[{{Proga}(2007)}]{Proga2007}
{Proga}, D. 2007, \apj, 661, 693, \dodoi{10.1086/515389}

\bibitem[{{Proga} {et~al.}(2008){Proga}, {Ostriker}, \& {Kurosawa}}]{Proga2008}
{Proga}, D., {Ostriker}, J.~P., \& {Kurosawa}, R. 2008, \apj, 676, 101,
  \dodoi{10.1086/527535}

\bibitem[{{Proga} \& {Waters}(2015)}]{Proga2015}
{Proga}, D., \& {Waters}, T. 2015, \apj, 804, 137,
  \dodoi{10.1088/0004-637X/804/2/137}

\bibitem[{{Reynolds}(1997)}]{Reynolds1997}
{Reynolds}, C.~S. 1997, \mnras, 286, 513, \dodoi{10.1093/mnras/286.3.513}

\bibitem[{{Ricci} {et~al.}(2017){Ricci}, {Trakhtenbrot}, {Koss}, {Ueda},
  {Schawinski}, {Oh}, {Lamperti}, {Mushotzky}, {Treister}, {Ho}, {Weigel},
  {Bauer}, {Paltani}, {Fabian}, {Xie}, \& {Gehrels}}]{Ricci2017}
{Ricci}, C., {Trakhtenbrot}, B., {Koss}, M.~J., {et~al.} 2017, \nat, 549, 488,
  \dodoi{10.1038/nature23906}

\bibitem[{{Shields} \& {Hamann}(1997)}]{Shields1997}
{Shields}, J.~C., \& {Hamann}, F. 1997, \apj, 481, 752, \dodoi{10.1086/304070}

\bibitem[{{Sim} {et~al.}(2012){Sim}, {Proga}, {Kurosawa}, {Long}, {Miller}, \&
  {Turner}}]{Sim2012}
{Sim}, S.~A., {Proga}, D., {Kurosawa}, R., {et~al.} 2012, \mnras, 426, 2859,
  \dodoi{10.1111/j.1365-2966.2012.21816.x}

\bibitem[{{Stone} {et~al.}(2020){Stone}, {Tomida}, {White}, \&
  {Felker}}]{Stone2020}
{Stone}, J.~M., {Tomida}, K., {White}, C.~J., \& {Felker}, K.~G. 2020, arXiv
  e-prints, arXiv:2005.06651.
\newblock \doarXiv{2005.06651}

\bibitem[{{Takeuchi} {et~al.}(2013){Takeuchi}, {Ohsuga}, \&
  {Mineshige}}]{Takeuchi2013}
{Takeuchi}, S., {Ohsuga}, K., \& {Mineshige}, S. 2013, \pasj, 65, 88,
  \dodoi{10.1093/pasj/65.4.88}

\bibitem[{{Waters} {et~al.}(2021){Waters}, {Proga}, \& {Dannen}}]{Waters2021}
{Waters}, T., {Proga}, D., \& {Dannen}, R. 2021, arXiv e-prints,
  arXiv:2101.09273.
\newblock \doarXiv{2101.09273}

\bibitem[{{Waters} {et~al.}(2017){Waters}, {Proga}, {Dannen}, \&
  {Kallman}}]{Waters2017}
{Waters}, T., {Proga}, D., {Dannen}, R., \& {Kallman}, T.~R. 2017, \mnras, 467,
  3160, \dodoi{10.1093/mnras/stx238}

\bibitem[{{Woods} {et~al.}(1996){Woods}, {Klein}, {Castor}, {McKee}, \&
  {Bell}}]{Woods1996}
{Woods}, D.~T., {Klein}, R.~I., {Castor}, J.~I., {McKee}, C.~F., \& {Bell},
  J.~B. 1996, \apj, 461, 767, \dodoi{10.1086/177101}

\bibitem[{{XRISM Science Team}(2020)}]{2020arXiv200304962X}
{XRISM Science Team}. 2020, arXiv e-prints, arXiv:2003.04962.
\newblock \doarXiv{2003.04962}

\end{thebibliography}
\end{document}